\documentclass{IEEEtran}
\usepackage{cite}
\usepackage{amsmath,amssymb,amsfonts,amsthm}
\usepackage{algpseudocode}
\usepackage[hyphens]{url}
\usepackage{algorithm}
\usepackage{algorithmicx}
\usepackage{graphicx}
\usepackage{textcomp}
\usepackage[dvipsnames]{xcolor}
\theoremstyle{definition}
\newtheorem{definition}{Definition}[section]

\theoremstyle{remark}

\DeclareMathOperator*{\argmax}{arg\,max}

\begin{document}

\title{Access Control for Distributed Ledgers in the Internet of Things: A Networking Approach}

\author{\thanks{This paper has been accepted for publication in the IEEE Internet of Things Journal. \\This research was funded in part by Science Foundation Ireland
		under grant 16/IA/4610 and in part by a grant from IOTA Foundation.}Andrew~Cullen\thanks{Andrew Cullen and Pietro Ferraro are with the Dyson School of Design Engineering at Imperial College London.}, Pietro~Ferraro,
  William~Sanders\thanks{William Sanders and Luigi Vigneri are with IOTA Foundation.}, Luigi~Vigneri, and
  Robert~Shorten\thanks{Robert Shorten is with the Dyson School of Design Engineering at Imperial College London and the School of Electrical \& Electronic Engineering at University College Dublin.}}%

\maketitle

\begin{abstract}
In the Internet of Things (IoT) domain, devices need
a platform to transact seamlessly without a trusted intermediary.
Although Distributed Ledger Technologies (DLTs) could provide
such a platform, blockchains, such as Bitcoin, were not designed
with IoT networks in mind, hence are often unsuitable for
such applications: they offer poor transaction throughput and
confirmation times, put stress on constrained computing and
storage resources, and require high transaction fees. In this
work, we consider a class of IoT-friendly DLTs based
on directed acyclic graphs, rather than a blockchain, and with
a reputation system in the place of Proof of Work (PoW).
However, without PoW, implementation of these DLTs requires
an access control algorithm to manage the rate at which nodes
can add new transactions to the ledger. We model the access
control problem and present an algorithm that is fair, efficient
and secure. Our algorithm represents a new design paradigm
for DLTs in which concepts from networking are applied to the DLT
setting for the first time. For example, our algorithm uses
distributed rate setting which is similar in nature to transmission
control used in the Internet. However, our solution features novel
adaptations to cope with the adversarial environment of DLTs
in which no individual agent can be trusted. Our algorithm
guarantees utilisation of resources, consistency, fairness, and
resilience against attackers. All of this is achieved efficiently and with
regard for the limitations of IoT devices. We perform extensive
simulations to validate these claims.
\end{abstract}


\section{Introduction}
\label{sec: introduction}
\begin{figure}
\centering
\includegraphics[width=\columnwidth]{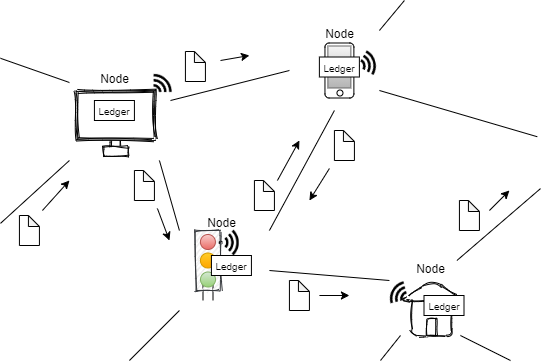}
\caption{{A section of a peer-to-peer IoT network sharing a distributed ledger. Nodes store a local copy of the ledger and share transactions with neighbouring nodes over bidirectional communication channels. Transactions can include exchanges of currency, IoT device data or any data required to be recorded on the ledger.}}
\label{fig: net model}
\end{figure}

\IEEEPARstart{D}{istributed} Ledger Technologies (DLTs) have recently received considerable attention for their potential applications in the Internet of Things (IoT) domain including facilitating machine-to-machine payment and securely exchanging data across platforms \cite{schiener2020data}. DLTs constitute an immutable record of data which is replicated over many nodes in a peer-to-peer network, providing enhanced transparency and integrity of data compared to their centralised counterparts. A section of a typical DLT network is illustrated in Figure \ref{fig: net model}.\newline

In the IoT setting, DLTs should be designed to facilitate high transaction throughput with low latency, and should be suitable for large-scale networks whilst retaining security and data integrity. These features are of paramount importance, particularly in applications involving the exchange of sensitive data, such as those from health monitoring devices \cite{tanwar2020blockchain}, or in mission-critical systems, such as real-time interactions between intelligent vehicles \cite{ferraro2018distributed}. Traditional DLTs, particularly blockchains, fall short of many of the aforementioned requirements for IoT-friendly DLT such as high throughput, low latency and network scalability. For this reason we consider a different class of DLTs in this work. Specifically, we are interested in ledgers with two fundamental features which differ from traditional DLTs: \emph{reputation-based Sybil protection}\footnote{We use the terms \emph{Sybil protection} and \emph{access control} interchangeably in this work. Their relationship is discussed in Section \ref{sec: related work}.}; and \emph{Directed Acyclic Graph} (DAG) ledger structure~\cite{wang2020sok}. As explained in more detail below, the former feature replaces the Proof of Work (PoW) most commonly employed in traditional DLTs, and the latter replaces the blockchain structure with a more general DAG.

{
\subsection{Motivation}	
The goal of this paper is to propose an access control mechanism for the aforementioned class of DLTs in order to guarantee full utilisation of network resources and fair access based on nodes' reputations. An alternative to PoW is sought because it is highly inefficient and offers unsatisfactory performance~\cite{vukolic2015quest}, and as recent attacks on the \emph{Nano} ledger have demonstrated\footnote{\url{https://www.coindesk.com/nanos-network-flooded-spam-nodes-out-of-sync}}, PoW is not secure in the context of IoT networks with limited resources. Our access control algorithm offers the first IoT-friendly alternative, allowing transaction throughput to be controlled across the network in a manner that is fair and resistant to manipulation by malicious actors. Our solution represents a new design paradigm for DLTs, permitting reputation-based access control to be integrated, in the place of PoW, into DAG-based ledgers for the first time.}
\subsection{Contributions}
The contributions of this work are as follows:
\begin{itemize}
    \setlength\itemsep{0.2em}
    \item We model the access control problem for a class of DAG-based DLTs and provide an accurate network model which takes into account limited buffer capacity and computational limitation of devices.
    \item We then present an access control algorithm for DAG-based distributed ledgers. The algorithm components include:
    \begin{itemize}
    \item[i)] A scheduling algorithm which ensures fair access for all nodes according to their reputation and prevents honest nodes from being adversely affected by malicious behaviour.
	\item[ii)] A rate setting algorithm, inspired by Transmission Control Protocol (TCP), which allows nodes to optimise their transaction issue rate in a decentralised manner.
	\item[iii)] A buffer management scheme to ensure that malicious actors can not cause buffers to overflow and compromise consistency for honest nodes.
	\end{itemize}
\item We provide extensive simulation results which demonstrate that the algorithm performs as intended, is robust to changes in the algorithm parameter choices, and is resilient against malicious actors who wish to tamper with the correct functioning of the algorithm.
\end{itemize}

{\textbf{Remark:} At the time of writing, implementation of our algorithm in IOTA's GoShimmer test network is underway~\cite{GoShimmer}. To the best of our knowledge, this algorithm is the first of its kind, although \emph{Nano} have recently announced primitive measures to prevent attacks on their PoW DAG which have similar motivation to our work\footnote{\url{https://senatusspqr.medium.com/nanos-latest-innovation-feeless-spam-resistance-f16130b13598}}.} {Note also that we do not give theoretical results, as their complexity and length would be outside the scope of this paper which seeks to introduce a novel solution to a problem of immense practical importance. Such mathematical analysis will be the subject of follow-up work.} \newline

The remainder of this paper is structured as follows: in Section \ref{sec: related work} we provide an overview of relevant material and prior art from both the DLT and broader networking domain. In Section \ref{sec: prob} we give a precise problem statement and state the requirements for our access control solution. In Section \ref{sec: model} we model the problem and provide the notation required to effectively describe and evaluate our solution. Section \ref{sec: algorithm} presents the access control algorithm in detail, while Section\ref{sec: evaluation} validates it and evaluates its efficacy through extensive simulations. Finally, in Section \ref{sec: conclusions} we summarise our results, provide conclusions and propose directions for future research.

\section{Background and Prior Art}
\label{sec: related work}
This work lies at the boundary of DLT and the broader networking literature, including topics such as TCP, Quality of Service (QoS), gossip protocols and many more. We begin with the necessary DLT backdrop which also serves to motivate our problem. As we have already mentioned, to the best of our knowledge, our networking approach to DLT access control is completely new, so our review of literature only covers related technologies rather than comparable algorithms. \newline

\subsection{Access Control for Distributed Ledgers}
Access control for DLTs refers to how nodes determine who gets to write new data to the ledger in a secure and distributed manner. This is also known as Sybil protection, because it prevents so called Sybil attacks in which an attacker creates multiple identities in order to gain an illegitimate advantage \cite{douceur2002sybil}. PoW access control, used in blockchains such as Bitcoin \cite{nakamoto2008bitcoin}, involves solving a computationally difficult puzzle to prove possession of computing resources to be allowed write to the ledger. PoW, however, consumes vast quantities of energy \cite{o2014bitcoin}, which is unacceptable from an environmental standpoint, unfeasible for IoT devices and inevitably concentrates computing power in the hands of those who can access specialised hardware and cheap energy. This is the case in the Bitcoin network \cite{kroll2013economics} where \emph{miners} select which transactions to include in blocks during busy periods (typically based on which offer the highest transaction fees), providing an intrinsic mechanism for filtering transactions and preventing congestion. \cite{huang2019towards} proposes a credit-based system for adapting the difficulty of the PoW for certain nodes which behave well, aiming to make PoW more suitable for IoT scenarios. \newline

The access control algorithm presented here, on the other hand, accommodates a more general \emph{reputation} model for Sybil protection mechanisms which do not require the use of computing resources, and are therefore more suitable for the IoT setting.
{
\begin{definition}[Reputation]
	Reputation is a quantity associated with each node's identity which is difficult to obtain and on which all nodes have consensus.
	\label{def: reputation}
\end{definition}
Reputation is an input to our access control algorithm. Proof of Stake (PoS) is an example of reputation-based access control in which reputation is the currency owned by a node.} In blockchains, PoS access control is typically implemented through some form of leader election in which a node or group of nodes becomes eligible to write a block each \emph{round} through some randomised process (e.g., \cite{androulaki2018hyperledger}, \cite{Ouroboros}, \cite{Casper}). Other examples of reputation-based access control include IOTA's \emph{mana} system \cite{popov2020coordicide}, delegated forms of PoS and {preconfigured node permissions as found in permissioned DLTs \cite{sujit2020pobt}.}

\subsection{DAG-Based Distributed Ledgers}
\begin{figure}[h]
\centering
\includegraphics[width=0.8\columnwidth]{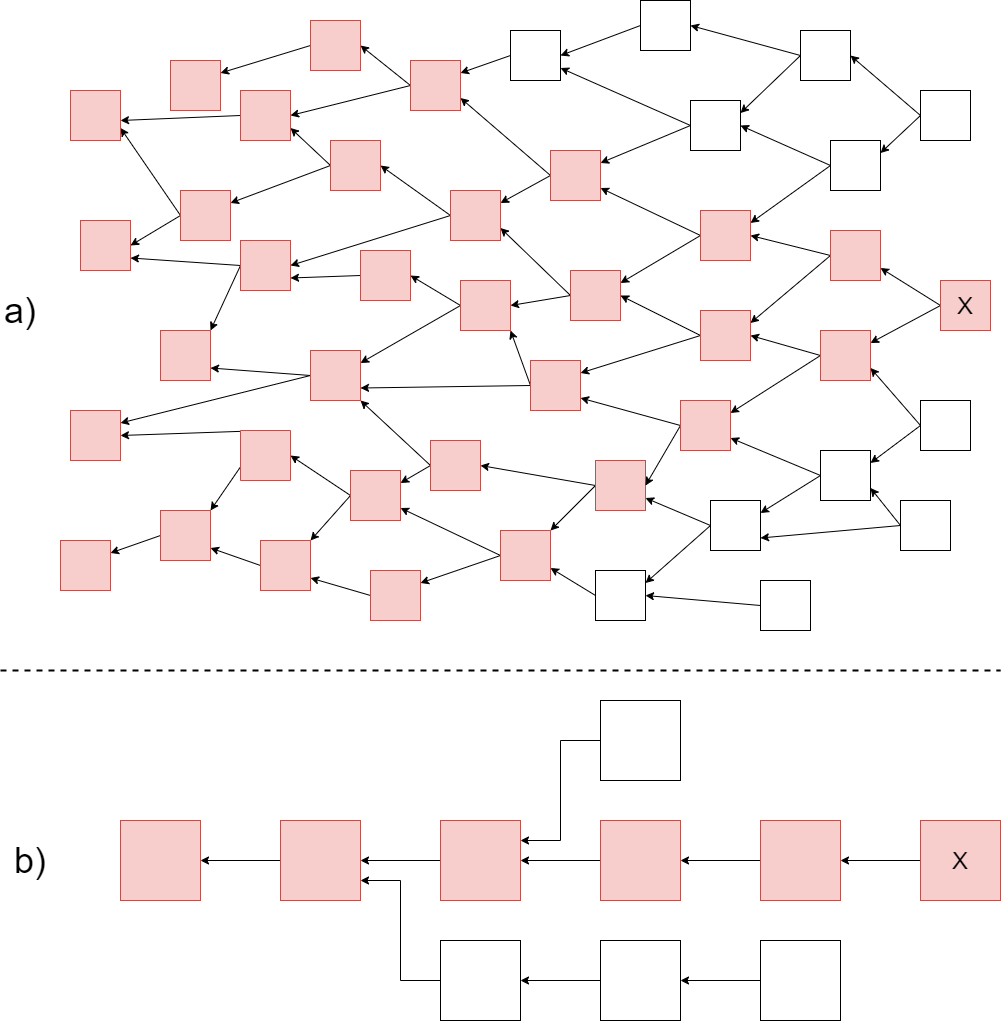}
\caption{a) a DAG ledger, and b) a blockchain ledger. Transactions/blocks of transactions are represented by squares and red transactions are those approved by transactions `X'. The white blocks become \emph{orphaned} in the blockchain, as only one chain can be approved by new transactions, but this is not the case in the DAG.}
\label{fig: DAG}
\end{figure}

Our interest in DAG-based DLTs, rather than blockchains, lies in their ability to accommodate high transaction throughput with low latency, and to support cooperative methods of consensus. Transactions can be added in blocks (a blockDAG \cite{lewenberg2015inclusive}), or as individual transactions \cite{popov2016tangle}. In this work we refer to the latter, but our framework is equally applicable to the former. In DAG-based ledgers, each new transaction can be cryptographically linked to more than one existing transaction, and many valid new transactions can be pointing to the same older transaction. The result is that many users of these ledgers can write transactions simultaneously, and therefore, there is no transaction throughput limit enforced by the ledger structure itself, as there must be in blockchains. Figure \ref{fig: DAG} compares a DAG and a blockchain. The precise details of how consensus is achieved in this setting is outside the scope of this paper and the interested reader may refer to \cite{cullen2020resilience} for recent work analysing aspects of security in DLTs of this kind. \newline

DAG-based ledgers traditionally rely on PoW for access control. {It is difficult to incorporate reputation-based access control such as PoS due to the lack of structure rounds for \emph{leader elections}.} \cite{bagaria2019prism} attempts to avoid this issue by using a so called structured DAG for their PoS ledger which imposes strict limitation on when and how transactions can be added to the ledger. Due to our focus on IoT scenarios, we do not want to impose any such additional constraints, and hence, the rate at which nodes can issue and disseminate transactions must be controlled by some other explicit mechanism. The access control algorithm we present here solves this problem by incorporating Sybil protection at the level of dissemination of transactions in the peer-to-peer (P2P) network, regulating transaction throughput on a per-node basis. This new paradigm requires concepts from networking which we review next.

\subsection{Networking Concepts}
From the domain of replicated databases, \cite{van2008efficient} presents a \emph{flow control} algorithm. The principle of what the authors of \cite{van2008efficient} aim to achieve is similar in nature to that of our access control algorithm, albeit in a trusted setting (without concern for adversarial behaviour). The flow control algorithm of \cite{van2008efficient} adaptively sets the update rate (transaction issue rate) of nodes using a TCP-like algorithm, but no measures are taken to defend against malicious nodes. A simple First In First Out (FIFO) scheduler is assumed to be used in \cite{van2008efficient}, and buffer overflows are used to signal congestion and reduce update rate. This presents an opportunity for malicious nodes to deflate the rate of others by setting their own rate too high. Additionally, trusted communication is used by nodes in \cite{van2008efficient} to achieve fairness in rates, which is also exploitable by malicious agents in the DLT setting. \newline

QoS in packet switched networks is related to our problem because we wish to fairly regulate flows of transactions from different sources. Classic examples of QoS architectures include Diffserv \cite{grossman2002new} and Intserv \cite{bernet2000framework}, which offer course and fine-grained QoS respectively. Both of these architectures rely on a backbone of trusted routers which are assumed to follow the protocol. Conversely, in DLT networks, no other individual node can be trusted to provide reliable information, making these classical architectures unsuitable. However, some of the core principles from these architectures are still applicable to DLTs, for example, the use of packet schedulers. A router employing a fair scheduling algorithm, in contrast to a simple FIFO scheduler, protects the flows of honest nodes from congestion caused by misbehaving flows \cite{nagle1987packet}. Fair schedulers have been proposed with varying levels of complexity, each trying to emulate generalised processor sharing as closely as possible. Weighted fair queuing \cite{demers1990analysis} provides a good approximation of GPS, but with significant computational overhead for routers, while simpler schedulers such as those based on Deficit Round Robin (DRR) \cite{shreedhar1996efficient}, \cite{macgregor2000deficits} provide lightweight and scalable alternatives. \newline

Another feature of QoS networks and of IP networks in general which is relevant to our access control problem is transmission control. TCP \cite{postel1981transmission} typically involves a distributed Additive Increase Multiplicative Decrease (AIMD) algorithm to set the transmission rate: nodes additively increase their transmission rate until congestion occurs (or is pre-empted), as signalled by some feedback from the network; and multiplicatively decrease their transmission rate in response to this congestion. In most forms of TCP, congestion is signalled when an acknowledgement is not received for a packet, and some other variants are based on Random Early Detection and require Explicit Congestion Notifications (ECNs) \cite{floyd1993random}. All of these AIMD algorithms still require feedback from other nodes, which leaves these protocols open to attack in the DLT setting, so none of them are applicable to our problem in their entirety.\newline

{\bf{Remark}:} The algorithm that we shall propose is designed to operate in adversarial environments (this is a baseline assumption under which DLTs are designed). Blockchain has been proposed for securing networks in certain settings (e.g., \cite{hu2020securing}), but the need for resilience to attacks is not typically considered in traditional networking applications such as those discussed above, which makes importing ideas from the traditional networking community difficult, and makes benchmarking our algorithm against similar work difficult, because it simply does not exist.

\section{Problem Statement}
\label{sec: prob}
We propose an access control algorithm for regulating transaction throughput, on a per-node basis, in a DAG-based DLT network with reputation-based Sybil protection. The goal of the algorithm is to allocate a portion of the network resources to each node proportional to their reputation and to prevent detrimental congestion. Transactions are issued by nodes and disseminated around the P2P network. Each node must validate all transactions, add them to its local copy of the ledger, and then run some consensus algorithm. We call these steps \emph{writing}. Writing is the bottleneck at which congestion can occur, and the goal of our algorithm is to allocate constrained resources fairly at this bottleneck. \newline

The specifics of writing will vary across DLT implementations and may even vary from node to node. {For example, in certain DLTs some nodes may do the most computationally heavy tasks while other limited nodes, such as IoT devices, perform lighter tasks while writing. Severely constrained devices may operate as \emph{light nodes}, relying on trusted \emph{full nodes} for a reliable view of the ledger\footnote{\url{trinity.iota.org/nodes}}.} \newline

Our access control algorithm seeks to maximise the rate of dissemination of transactions, subject to the writing bottleneck, while minimising delays. The algorithm must also meet the requirements listed below. These requirements are described at a very high level here, and defined more precisely before presenting our access control algorithm in Section \ref{sec: algorithm}.
\begin{itemize}
    \item \emph{Consistency}: If a transaction issued by an honest node\footnote{A node that follows the proposed protocol.} is written by one honest node, it should eventually be written by all honest nodes.
    \item \emph{Fairness in dissemination rate}: The dissemination rate of each node should be allocated fairly according to the node's reputation.
    \item \emph{Fairness in latency}: For a given dissemination rate, relative to the node's reputation, a node's transactions should experience similar latency.
    \item \emph{Security}: Malicious nodes\footnote{A node that arbitrarily deviates from the proposed protocol.} should be unable to interfere with any of the above requirements.
\end{itemize}

{The physical limits of devices and the particular consensus algorithm employed determine the rate of the writing bottleneck. This in turn determines the maximum performance of the network (transactions per second, confirmation time, etc.), and our access control algorithm ensures that this maximum performance can be reached.}

\section{Model and Notations}
\label{sec: model}

\begin{figure*}[ht]
\includegraphics[width=0.8\textwidth]{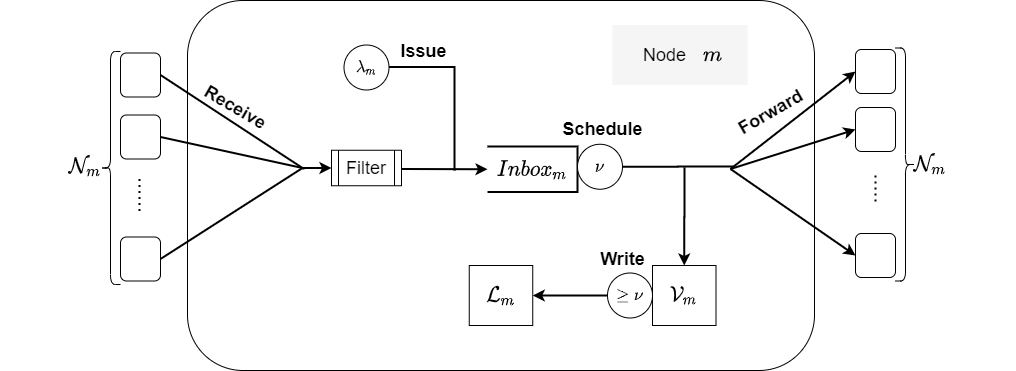}
\centering
\caption{Model for a node $m$, indicating the actions available to each node, namely, receiving, issuing, scheduling, writing and forwarding transactions. $\lambda_m$ denotes node $m$'s transaction issue rate, $\nu$ denotes its maximum scheduling rate, and with $\geq \nu$ denotes its writing rate which must be at least $\nu$.}
\label{fig: node model}
\end{figure*}
{The model introduced in this section is illustrated in Figure \ref{fig: node model}, and the associated notation is summarised in Table \ref{tab: notation} at the end of this section. The ledger is distributed over a set of nodes $\mathcal{M}$ in a P2P network, where a node $m$ in $\mathcal{M}$ has a set of neighbours $\mathcal{N}_m \subset \mathcal{M}$ with which it communicates directly.  The subset of nodes which correctly follow the protocol, referred to as honest nodes, is denoted $\mathcal{M}^*$. The reputation distribution over the nodes (see Definition \ref{def: reputation}) is denoted $rep$, where $rep_m$ denotes the reputation of node $m$. In the experiments presented here, $rep$ is assumed not to vary with time.} \newline

{Transactions are cryptographically signed, which links them to the identity of their issuer. The set of transactions that are visible to node $m$ (either issued by node $m$ itself or received from neighbours) is denoted by $\mathcal{V}_m$. Each node additionally adds a subset of these visible transactions to their local ledger, $\mathcal{L}_m \subseteq \mathcal{V}_m$. Consensus must be reached for a transaction to be added to $\mathcal{L}_m$, so a decision must made for each transaction in $\mathcal{V}_m$ whether it should be written to $\mathcal{L}_m$.
}

\begin{definition}[Disseminated transaction]\label{def: dissemrate}
We say that a transaction is \emph{disseminated} when it has been received by all honest nodes, $\mathcal{M}^*$. The set of disseminated transactions, $\mathcal{D}$, is defined as follows:
\begin{equation}
\mathcal{D} = \bigcap_{m \in \mathcal{M}^*}\mathcal{V}_m \\
\label{eq: dissem def}
\end{equation}
and $\mathcal{D}_i$ denotes the subset of transactions in $\mathcal{D}$ which were issued by node $i$.

\end{definition}

{Nodes in a given DLT have some constrained resource (e.g., computation or storage) that limits the rate at which they can process incoming transactions.} The \emph{writing work} of a transaction $tx$ is the work required from this constrained resource at each node in order to reach consensus and decide if $tx$ should be added to a node's local ledger, $\mathcal{L}_m$. We assume the expected writing work for $tx$, denoted by $|tx|$, to be known in advance as it depends on known information such as transaction payload type or size. Note that this modeling choice permits to have a flexible algorithm where specific class of transactions can be prioritised, if needed. The expected writing work required to write the transactions in a set $A$ is denoted by $W(A)$, i.e.: 
\begin{equation}
W(A) = \sum_{tx \in A}|tx|
\end{equation}

and \emph{writing power} corresponds to the rate at which this writing work is done. As writing power is the limited resource in our network model, the rate of dissemination of transactions must be measured in terms of the power required to write them.
\begin{definition}[Dissemination rate]
The \emph{dissemination rate}, $DR$, is the rate of dissemination of transactions, weighted by their work. This dissemination rate and the dissemination rate of node $i$'s transactions, respectively, are defined as follows:
\begin{eqnarray}
DR &=& \frac{\Delta W(\mathcal{D})}{\Delta t} \\
DR_i &=& \frac{\Delta W(\mathcal{D}_i)}{\Delta t}
\end{eqnarray}
\end{definition}
Where $\Delta t$ is the time window over which we measure the dissemination rate. \newline

Another important quantity for evaluating the performance of our access control algorithm is latency, which is defined as follow:
\begin{definition}[Latency]\label{def: latency}
The latency of a transaction is the time from when the transaction is issued to when it is added to $\mathcal{D}$. In other words, latency is the random variable of the time it takes for a transaction to reach all honest nodes after it is issued.
\end{definition}

Note that in order to achieve a consistent distributed ledger, all nodes must possess some minimum writing power to ensure that they can write transactions sufficiently quickly to keep up with the network's dissemination rate $DR$. For this reason we enforce a global writing power, $\nu$, which all nodes must be able to achieve. \newline

The actions taken by nodes are highlighted in boldface on the node model diagram in Figure \ref{fig: node model}, and described as follows:

\subsection{Receive}
We assume reliable communication channels\footnote{Point-to-point connections are handled with TCP on a separate network layer, and hence can be considered reliable.}, hence a node $m$ must receive all transactions sent by its neighbours $\mathcal{N}_m$. These transactions are filtered at this point to remove duplicates and invalid transactions. A transaction can be considered invalid depending on the specific protocol specifications: at a high level, this filtering criterion concerns signature and timestamp validation, and protection against denial of service attacks \cite{attias2020preventing}. Filtered transactions are added to node $m$'s inbox buffer, $Inbox_m$.

\subsection{Issue}
Nodes can additionally issue their own transactions, and these transactions are also added to the issuing node $m$'s inbox buffer, $Inbox_m$. The rate at which node $m$ issues transactions is denoted by $\lambda_m$, and this is controlled by node $m$ using a rate setting algorithm. If all nodes wished to have transactions written at all times, then a fair allocation of the writing power would permit each node to have an assured issue rate, $\Tilde{\lambda}_m$, defined as:
\begin{equation}
    \Tilde{\lambda}_m = \frac{\nu \cdot rep_m}{\sum_{i \in \mathcal{M}}{rep_i}}.
\end{equation}
{with units of work per second. Assuming a fair allocation of writing resources in the network can be achieved, node $m$ can issue transactions at a rate less than or equal to $\Tilde{\lambda}_m$, safe in the knowledge that these transactions will be written by all nodes without causing backlogs and delays, regardless of what rate other nodes issue transactions at. If a node wishes to issue transactions at a rate greater that $\Tilde{\lambda}_m$, this must be done taking the issue rate of other nodes into account so as to avoid excessive congestion. This latter observation motivates the need for the \emph{rate setting} component of our access control algorithm to allow nodes to effectively use excess capacity.}

To capture the varying demand across nodes to issue transactions, we define four modes of operation for nodes issuing transactions:

\begin{definition}[Inactive]\label{def: inactive}
A node is said to be in \emph{inactive} mode if it is not issuing any transactions, i.e., $\lambda_m=0$.
\end{definition} 
\begin{definition}[Content]\label{def: content}
A node is said to be in \emph{content} mode if it is issuing transactions at a fixed rate $\lambda_m\leq\Tilde{\lambda}_m$. This is modelled as a Poisson process with rate parameter $\lambda_{m}$, which is a standard model for arrival processes.
\end{definition}
\begin{definition}[Best-effort]\label{def: best-effort}
A node is said to be in \emph{best-effort} mode if it is issuing transactions at the highest rate possible under the current traffic conditions, without causing excessive congestion. This requires a node to use the rate setting algorithm, outlined in Section \ref{sec: algorithm}, to utilise unused network resources and adaptively set $\lambda_m>\Tilde{\lambda}_m$. We assume that a leaky bucket regulator with rate $\lambda_m$ is used to achieve the set issue rate i.e.\ the rate is deterministic, rather than Poisson.
\end{definition}
\begin{definition}[Malicious]\label{def: malicious}
A node is said to be in \emph{malicious} mode if it is issuing transactions at a rate  $\lambda_m\gg\Tilde{\lambda}_m$, without concern for the congestion caused. We assume that a malicious node immediately writes and forwards its own transactions rather than including them in the scheduling steps.
\end{definition}

\subsection{Schedule}
Transactions issued by node $m$ itself and those received from neighbours are all added to $Inbox_m$, as described above. $Inbox_m(i)$ denotes the transactions in $Inbox_m$ issued by node $i$. Transactions from $Inbox_m$ are then scheduled, added to the set of visible transactions $\mathcal{V}_m$, and forwarded to neighbours. A fair scheduling algorithm \cite{nagle1987packet} should be used to ensure that malicious agents issuing transactions at an excessive rate can not delay the transactions of honest nodes at the inbox buffers. This scheduling algorithm is discussed further in Section \ref{sec: algorithm}. \newline

The scheduling process is deterministic, with rate $\nu$, when there are transactions in the inbox to be scheduled. Note that this deterministic scheduling rate is in units of writing power and is imposed to ensure that the visible set of transactions can be written by all nodes having at least the minimum writing power $\nu$.

\subsection{Write}
When a transaction has been scheduled, it is added to $\mathcal{V}_m$. The transaction is not yet considered part of the ledger $\mathcal{L}_m$ at this point. Rather, the transaction must still satisfy the consensus rules of the DLT. The consensus protocol can vary between DLTs. For example, something comparable to the \emph{longest chain rule} of Bitcoin may be applied to DAGs. The interested reader can refer to \cite{cullen2020resilience} for further details on this kind of consensus algorithm in DAGs. Alternatively, or indeed additionally, a voting algorithm such as \cite{popov2019fpc}, \cite{muller2020fast} can be used to achieve consensus in the presence of conflicting \emph{branches} of the DAG. \newline

The complexity of the aforementioned consensus protocols increases with the number of transactions involved, and varies with the type of transactions involved (for example, transactions containing sensor measurement data may be cheaper to reach consensus on than those transferring currency between accounts). The limit $\nu$ is a parameter of the DLT which is configured to ensure that any node with some minimum writing power can participate in the consensus.

\subsection{Forward}
After a transaction is scheduled by node $m$, it is forwarded to all neighbours $\mathcal{N}_m$ except for the neighbours from which the transaction has already been received (these are the only nodes that $m$ can be sure already have this transaction). This is known as \emph{flooding}, and while it is a highly inefficient use of communication resources, alternative approaches based on gossiping can only provide probabilistic guarantees of transaction dissemination and the associated latency. For the sake of robustness and simplicity of analysis, we do not consider any such optimisation in this work.

\begin{table}[ht]
\caption{Notation for node and network model.}
\centering
 \begin{tabular}{c|l}

 $\mathcal{M}$      & set of all nodes in the network\\
 $\mathcal{M}^*$    & set of all honest nodes \\
 $\mathcal{N}_m$    & set of nodes that are neighbours of node $m$ \\ 
 $rep_m$            & reputation of node $m$ \\
 $\mathcal{V}_m$    & set of transactions visible to node $m$ \\ 
 $\mathcal{L}_m$    & set of finalised transactions in node $m$'s ledger \\
 $\mathcal{D}$		& set of all disseminated transactions \\
 $\mathcal{D}_i$	& set of disseminated transactions issued by node $i$ \\
 $DR$	& dissemination rate (all transactions) \\
 $DR_i$	& dissemination rate (transactions issued by node $i$) \\
 $\nu$              & global transaction writing power \\
 $\lambda_m$        & issue rate of node $m$ \\
 $\Tilde{\lambda}_m$ & assured issue rate of node $m$ \\
 \hline\hline
\end{tabular}
\label{tab: notation}
\end{table}

\subsection{Definition of Requirements}
We now provide more precise definitions for each of the requirements stated in Section \ref{sec: prob}.

\begin{definition}[Consistency]\label{def: consistency}
	Consider a finite time window $w\in\mathbf{R}^+$, and a finite offset $h\in\mathbf{R}^+$. At time $t+h$, if all transactions added to $\mathcal{V}_m$ for any $m$ within time $[t - w, t]$ are in $\mathcal{D}$, the access control algorithm of this network is said to satisfy the consistency requirement.
\end{definition}

{The interpretation of this consistency requirement all nodes must eventually receive all transactions, which is essential if consensus on the ledger is to be achieved.}

\begin{definition}[Fairness in dissemination rate]\label{def: fairdissem}
	An access control algorithm satisfies the \emph{fairness in dissemination rate} requirement if allocation of dissemination rate among nodes is max-min fair, weighted by each node's reputation. An allocation is max-min fair if an increase in any node's dissemination rate decreases the dissemination rate of another node $m$ with equal or smaller \emph{reputation-scaled} dissemination rate, $DR_m / rep_m$.
\end{definition}

This fairness in dissemination rate requirement ensures that network resources are allocated to nodes based on their reputation so that each node gets a fair access to the ledger, and hence a fair vote.

\begin{definition}[Fairness in latency]\label{def: fairlat}
	We say that an access control algorithm satisfies the \emph{fairness in latency} requirement if the expected\footnote{This property may also be defined in terms of maximum latency, rather than expected latency, depending on the requirements of the specific ledger.} latency of a node's transactions is independent of its reputation, and increases with its reputation-scaled dissemination rate.
\end{definition}

This fairness in latency requirement ensures that transactions belonging to particular nodes do not experience excessive delays. This is a soft requirement, and approximate fairness in latency is sufficient.

\begin{definition}[Security]\label{def: security}
	An access control algorithm satisfies the \emph{security} requirement if the requirements defined in Definitions \ref{def: consistency}--\ref{def: fairlat} are still satisfied in the presence of malicious actors.
\end{definition}

This final requirement of security is essential for the DLT environment in which some nodes (malicious actors) may decide to try to gain an unfair advantage by deviating from the protocol. The security requirement ensures that this can not happen.

\section{Access Control Algorithm}
\label{sec: algorithm}
The relevant notation and definitions are now in place, and we can now present our solution which consists of three core components, namely scheduling, rate setting and buffer management. \newline

\emph{Scheduling:} The scheduling component aims to ensure that transactions issued by honest nodes do not experience delays due to congestion caused by dishonest nodes. To this end, in the presence of congestion, transactions should be scheduled at a rate proportional to the reputation of the node that issued the transactions.

\emph{Rate setting:} The rate setting component seeks to allow \emph{best-effort} nodes (see Section \ref{sec: model}) to issue at a rate above their assured rate, $\Tilde{\lambda}_m$, without causing excessive congestion and large delays which could cause a violation of the consistency requirement.

\emph{Buffer management:} The buffer management component decides when to drop transactions to protect honest nodes' transactions from being dropped due to congestion caused by malicious nodes. Under normal network operation, the rate setting component should prevent the need for any dropped transactions, but in the event of a malicious agent issuing transactions at an excessively high rate, buffer management can ensure that only this malicious node's transactions are dropped and the buffer does not reach its physical capacity.

\subsection{Scheduling}
Nodes in our setting are capable of more complex and customised behaviour than a typical router in a packet-switched network, but our scheduler must still be efficient and scalable due to the potentially large number of nodes requiring differentiated treatment. It is estimated that over 10,000 nodes operate on the Bitcoin network\footnote{\url{https://bitnodes.io/}.}, and we expect that an even greater number of nodes are likely to be present in the IoT setting. We therefore adopt an efficient and scalable scheduler based on Deficit Round Robin (DRR), with modifications to deal with particular features of a DLT network, namely high variance in reputation among nodes and potentially bursty traffic. \cite{shreedhar1996efficient}. {The standard Linux implementation of the DRR-based scheduler used in \cite{hoeiland2018flow} permits up to 65,535 separate queues, which demonstrates the scalability of these methods.}  \newline

\begin{figure}
\includegraphics[width=\columnwidth]{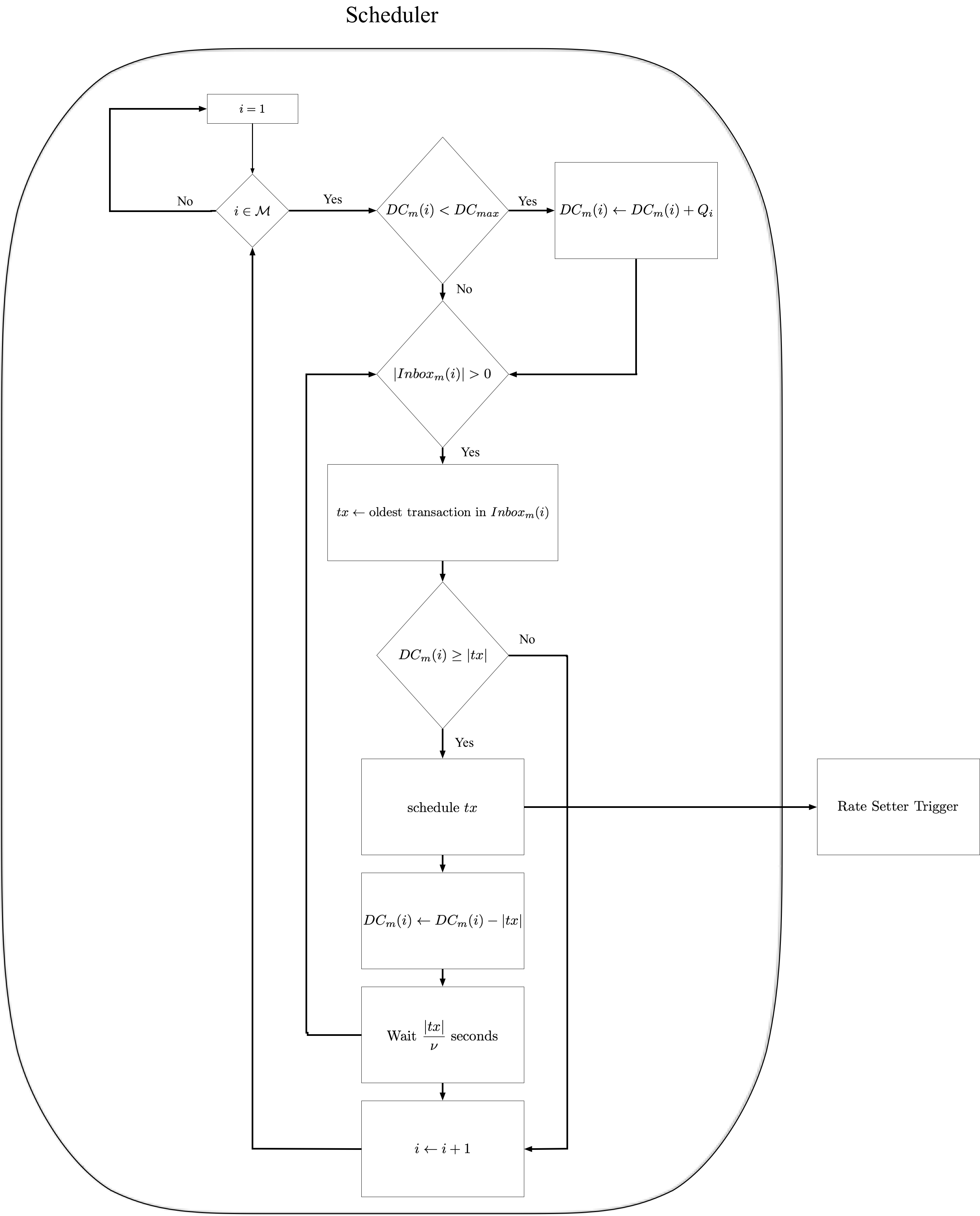}
\centering
\caption{{Flowchart of Algorithm 1. Notice that whenever a transaction is scheduled Algorithm 2 is triggered.}}
\label{fig: Alg1}
\end{figure}

{DRR-based scheduling algorithms are very simple: each flow of packets (transactions) is visited in a round robin cycle, and deficit is assigned to the flow. Deficit can be thought of as credits to schedule packets, where sufficient credits must be accrued by a flow in order to have a packet scheduled. Our scheduling algorithm, DRR$-$, is presented in Algorithm \ref{alg: drr-} and illustrated in the flowchart in Figure \ref{fig: Alg1}. Node $m$ maintains a deficit counter, $DC_m(i)$, for each node $i$ in $\mathcal{M}$. Each node in $\mathcal{M}$ is considered in a round robin cycle, one after another. Regardless of whether $Inbox_m(i)$ has any transactions in it, $DC_m(i)$ is incremented by a quantum, $Q_i$, which is proportional to $rep_i$, up to a maximum $DC_{max}$. The unit of deficit here is that of writing work, and deficit $|tx|$ must be spent from $DC_m(i)$ in order to schedule a transaction $tx$ from $Inbox_m(i)$. Transactions in $Inbox_m(i)$ for each $i$ are scheduled in FIFO order. The parameter $DC_{max}$ should be chosen to be higher than the maximum work required to write a single transaction, and such that $Q_i \ll DC_{max}$ for all nodes.} \newline

In standard DRR \cite{shreedhar1996efficient}, a flow must be backlogged (packets from this flow must be waiting in the queue) in order to gain deficit. This feature of DRR presents problems for bursty traffic, because queues may periodically empty between bursts. This is an issue in DLT networks because transactions must traverse multihop paths in the P2P network and the dynamics of these paths result in bursty arrivals at nodes' inbox queues. In our setting, higher reputation nodes can issue at a higher rate, meaning that queues are backlogged with their transactions more often even when the arrival of their transactions is bursty. Conversely, low reputation nodes issue at a lower rate, and queues may be completely emptied of their transactions between bursts. This gives higher reputation nodes an advantage in the standard DRR scheduler. DRR++ is a modification of the standard DRR scheduler designed to ensure low delays for so-called latency-critical flows in the presence of bursty arrivals \cite{macgregor2000deficits}. DRR++ performs well for a small portion of latency-critical flows but becomes more comparable to standard DRR when all flows are deemed latency-critical, as we require in the DLT setting. \newline

DRR$-$, on the other hand, can accommodate bursty traffic from all nodes. The principle behind DRR++ is essentially to allow latency-critical flows to go into negative deficit so that they can be scheduled rapidly. Our approach instead allows flows to gain deficit up to some limit, even when the flow is not backlogged, saving rather than going in to debt. {This allows bursty traffic while maintaining efficiency because nodes with maximum saved deficit (indicating that they are inactive) do not need to be visited by the scheduler.} The scheduler presented in \cite{menth2018deficit} also uses a concept of deficit savings to accommodate bursty flows. However, \cite{menth2018deficit} does not permit differentiated treatment of flows based on reputation and their analysis focuses primarily on active queue management (AQM) to prevent bufferbloat \cite{gettys2011bufferbloat}.

\begin{algorithm}[h!]
\caption{DRR$-$ Scheduler}\label{alg: drr-}
\begin{algorithmic}[1]
\Statex \emph{Repeat for $i \in \mathcal{M}$ in a round robin cycle:}
\If{$DC_m(i) < DC_{max}$} 
	\State $DC_m(i) \gets DC_m(i) + Q_i$
\EndIf
\While{$\left|Inbox_m(i) \right|>0$}
    \State $tx \gets$ oldest transaction in $Inbox_m(i)$
    \If{$DC_m(i)\geq \left|tx\right|$}
		\State Schedule $tx$
    	\State $DC_m(i) \gets DC_m(i) - |tx|$
		\State Wait $\frac{|tx|}{\nu}$ seconds
    \Else
    	\State break
	\EndIf
\EndWhile

\end{algorithmic}
\end{algorithm}

\begin{table}[ht]
\caption{Scheduling algorithm parameters.}
\centering
\begin{tabular}{c|l}
 $DC_j(i)$          & deficit counter for transactions in $Inbox_j(i)$ \\
 $Q_i$              & quantum added to $DC_j(i)$, $\forall j$ in each round ($\propto rep_i$) \\
 $DC_{max}$			& maximum deficit for an empty queue \\
 \hline\hline
\end{tabular}
\label{tab: sched notation}
\end{table}

\subsection{Rate Setting}
If all nodes always had transactions to issue, the problem of rate setting would be very straightforward: nodes could simply operate in \emph{content} mode, at a fixed, assured rate, $\Tilde{\lambda}_m$ (see Definition \ref{def: content}). The scheduling algorithm ensures that this rate is enforceable and that increasing delays or dropped transactions are only experienced by misbehaving nodes. However, it is highly unlikely that all nodes will always have transactions to issue, and we would like \emph{best-effort} nodes to better utilise network resources, without causing excessive congestion and violating requirements. \newline

Our rate setting algorithm, for \emph{best-effort} nodes, is inspired by TCP --- each node uses AIMD (see Section \ref{sec: related work}) rules to update their issue rate in response to congestion events \cite{corless2016aimd}. However, in the trustless DLT setting, the traditional means of responding to congestion is compromised. For example, malicious nodes could attempt to deflate the issue rate of their neighbours by not sending acknowledgements, or sending illegitimate congestion notifications. We recall, however, that in distributed ledgers, all transaction traffic passes through all nodes, contrary to traffic typically found in packet switched networks and other traditional network architectures. Under these conditions, local congestion at a node indicates congestion elsewhere in the network. This observation is crucial, as it presents an opportunity for an access control algorithm based entirely on local traffic and without the need for additional, potentially corruptible, messages between nodes. \newline

Recall that when a node $m$ issues a transaction, it is added to its inbox buffer to be scheduled. Node $m$'s own transactions in its inbox, $Inbox_m(m)$, are then scheduled at a rate which depends on the other traffic present in the buffer. We observe that the length of $Inbox_m(m)$ gives an estimate of congestion in node $m$'s traffic, not only at its own inbox buffer but at $Inbox_i(m)$ for all nodes $i$ in $\mathcal{M}^*$, within some network delay. \newline

\begin{figure}
\includegraphics[width=\columnwidth]{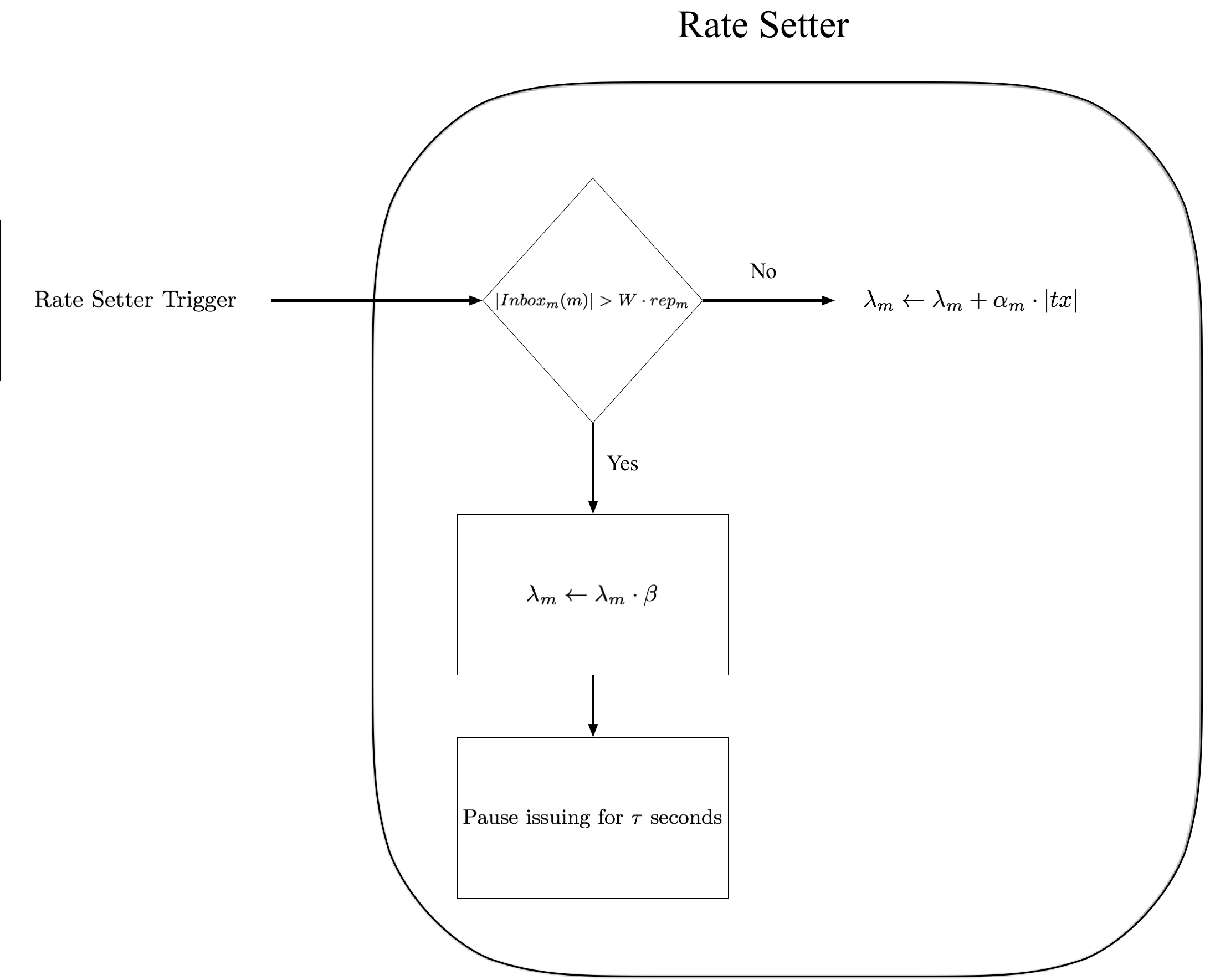}
\centering
\caption{{Flowchart of Algorithm 2.}}
\label{fig: Alg2}
\end{figure}

{Algorithm \ref{alg: AIMD} outlines the AIMD rules used by each node to set their issue rate, and the parameters of the rate setting algorithm are outlined in Table \ref{tab: rs notation}. The algorithm is also illustrated in the flowchart in Figure \ref{fig: Alg2}.} Each node sets their own local additive-increase parameter based on the global increase rate $A$, and their reputation. Specifically, each node sets their local increase parameter as follows:
\begin{equation}
\alpha_m \gets A \cdot \frac{rep_m}{\sum_{i}{rep_i}}
\end{equation}
An appropriate choice of $A$ ensures a conservative global increase rate which does not cause problems even when many nodes increase their rate simultaneously. Updates are made to the issue rate each time a transactions is scheduled, at a rate proportional to the writing work of the transaction, $|tx|$, which allows the rate setter to accommodate variable transaction types and sizes. Nodes wait $\tau$ seconds after a multiplicative decrease, during which there are no further updates made, to allow the reduced rate to take effect and prevent multiple successive decreases. Waiting after decreases is common in implementations of AIMD algorithms, such as sliding window flow control in  TCP \cite{postel1981transmission}. The rate is updated each time a transaction is scheduled which at rate $\nu$ when the inbox is not empty. At each update, node $m$ checks the work $|Inbox_m(m)|$ and responds with a multiplicative decrease if this is above a threshold, $W_m$, which is proportional to $rep_m$. If $|Inbox_m(m)|$ is below this threshold, $m$'s issue rate is incremented by its local increase parameter $\alpha_m$. $|Inbox_m(m)|$ is measured as an exponential moving average with samples taken each time a transaction is scheduled.

\begin{table}[ht]
\caption{Rate setting algorithm parameters.}
\centering
\begin{tabular}{c|l}
 $A$      		& global additive increase parameter \\
 $\beta$        	& global multiplicative decrease parameter \\
 $\tau$         	& wait time parameter \\
 $W$             & inbox work threshold \\
 \hline\hline
\end{tabular}
\label{tab: rs notation}
\end{table}

\begin{algorithm}[h!]
\caption{AIMD Rate Setter (Best-effort Mode)}\label{alg: AIMD}
\begin{algorithmic}[1]
\Statex \emph{Repeat each time a transaction $tx$ is scheduled:}
\If{$\left|Inbox_{m}(m)\right|> W \cdot rep_m$} \label{line: backoff conditions}
    \State $\lambda_{m} \gets \lambda_{m} \cdot \beta$ \label{line: decrease}
    \State Pause issuing and rate setting for $\tau$ seconds
\Else
    \State $\lambda_{m} \gets \lambda_{m} + \alpha_{m} \cdot |tx|$
\EndIf
\end{algorithmic}
\end{algorithm}


\subsection{Buffer Management}
The inbox buffers do not have infinite capacity, and buffer capacity can be particularly limited in the case of IoT devices. Even if the capacity of buffers could be made arbitrarily large, an excessively full buffer would result in large delays \cite{gettys2011bufferbloat}. Our buffer management seeks to drop transactions fairly (with respect to issuing node's reputation) whenever the buffer exceeds a certain size. The objective of our buffer management differs from than that of typical active queue management (AQM) systems. AQM is generally used to regularly drop packets and generate explicit congestion notifications (ECNs), forming a key component of how congestion is detected for rate setting. An early example of this kind of AQM can be found in \cite{floyd1993random}, and it is also a component of the schedulers discussed above  \cite{menth2018deficit}, \cite{hoeiland2018flow}. However, dropped transactions or ECNs are not used to detect congestion in our setting, so our buffer management does not play a direct role in rate setting. Rather, buffer management plays the important role of ensuring that malicious nodes that do not abide by the rate setting rules can not fill the buffers and cause honest transactions to be dropped. Our rate setting algorithm prevents excessive congestion under normal operation and buffer management should only take effect in the presence of malicious behaviour. \newline

The simple buffer management rule is stated in Algorithm \ref{alg: bm}. This buffer management is equivalent to the \emph{Longest Queue Drop} scheme proposed in \cite{suter1998design}. The parameter $W_{max}$ is the maximum work in the inbox buffer after which transactions should be dropped. We assume that each transaction has some maximum size in memory and requires some minimum work, so $W_{max}$ bounds the memory needed. If the buffer memory is specified to be larger than this worst case memory, the buffer management will prevent the buffer from overflowing. \newline

If the work in the buffer exceeds $W_{max}$, the node with the greatest amount of work in the buffer, relative to its reputation, is identified, and the first received transaction is dropped, as outlined in Algorithm \ref{alg: bm}. This buffer management strategy ensures that consistency is preserved for honest nodes, because the work from honest nodes in the buffer should remain modest and hence their honest transactions will not be dropped --- the scheduler ensures that this is the case as it will continue to schedule fairly without regard for large influxes of transactions from malicious nodes.

\begin{algorithm}[h!]
\caption{Buffer Management}\label{alg: bm}
\begin{algorithmic}[1]
\While{$|Inbox_m|>W_{max}$}
	\State $d \gets \argmax_{i \in \mathcal{M}} {\frac{|Inbox_m(i)|}{rep_i}}$
	\State Drop transaction from head of $Inbox_m(d)$
\EndWhile
\end{algorithmic}
\end{algorithm}

\section{Evaluation}
\label{sec: evaluation}
We now present extensive simulations to evaluate the efficacy of our approach. The main focus of our evaluation is to verify that the requirements have been met, as these guarantee that the resources available to the nodes are optimally utilised both fairly and securely. To this end, we provide simulations for both honest and malicious environments, then demonstrate the robustness of our solution to parameter choices. After that, we demonstrate how our algorithm can accommodate different node types and their differing usage of the ledger in IoT settings. Finally, we benchmark our work against a PoW, which represents the current state of the art access control solution for DAG-based DLTs. \newline

{In terms of evaluating performance, our results show that we can achieve close to 100\% utilisation of node resources up to the limit $\nu$ set to ensure all nodes can keep up with the rate of newly issued transactions. The number of transactions per second that a DLT employing this access control can handle will therefore depend on the resources available to nodes and the consensus algorithm employed which together will determine the achievable rate $\nu$.} \newline

The results in this section are produced with a Python simulator for DAG-based distributed ledgers\footnote{Source code available at {\url{https://github.com/cyberphysic4l/DLTCongestionControl}.}}. We consider a network of 50 nodes, each storing a copy of the ledger, and each with 4 randomly selected neighbours. The mean propagation delay of each communication channel between neighbours is chosen uniformly at random between 50 ms and 150 ms, and the delay for each transaction on these channels is normally distributed around this average with standard deviation 20 ms. Node reputation is computed according to real data, that is, the number of transactions issued by each account in the IOTA network and follows a Zipf distribution\footnote{Wealth has also been shown to follow similar distributions, so this model is also well suited to reputation systems derived from wealth, i.e., PoS \cite{jones2015pareto}.} with exponent 0.9. Simulation results are averaged over 20 monte carlo simulations, each 180 seconds of simulation time. We assume that each node has buffer capacity greater than the parameter $W_{max}$ specified for the buffer management, so no buffer overflows occur in an honest environment.

\subsection{Honest Environment}\label{subsec: honest}
\begin{figure}[ht]
\centering
\includegraphics[width=\columnwidth]{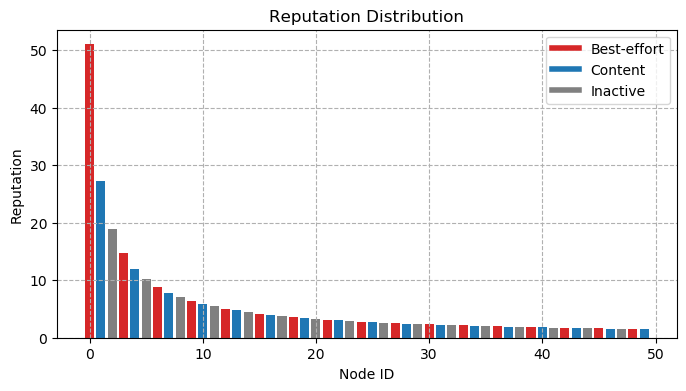}
\caption{Reputation distribution follows a Zipf distribution with exponent 0.9. Nodes are Content, Best-effort, or Inactive as indicated by each bar's colour.}
\label{fig: repdist 1}
\end{figure}

The first set of simulations is in an honest environment, where each node is operating in one of the three honest modes: inactive, content, or best-effort (see Definitions \ref{def: inactive}--\ref{def: best-effort}). We evaluate whether dissemination rate is maximised, latency is minimised, and the first three requirements are met, namely consistency, fairness in dissemination rate and fairness in latency (see Definitions \ref{def: consistency}--\ref{def: fairlat}). The distribution of reputation and operating mode is illustrated in Figure \ref{fig: repdist 1}. The global transaction writing rate is $\nu=50$ units of work per second and each transaction requires one unit of work in these initial simulations, i.e., $|tx|=1$ for all transactions $tx$. The parameters of the access control algorithm for this set of simulations, given in Table \ref{tab: sim params 1}, are chosen experimentally.

\begin{table}[ht]
\caption{Access control algorithm parameters.}
\centering
\begin{tabular}{c c c || c c c c|| c ||}
\multicolumn{3}{c||}{\textbf{Scheduler}} & \multicolumn{4}{c||} {\textbf{Rate Setter}} & \textbf{Buffer Man.} \\
 $\nu$ & $Q_i$ & $DC_{max}$ & $A$ & $\beta$ & $\tau$ & $W$ & $W_{max}$ \\
 \hline
 $50$ & $\frac{rep_i}{\sum{rep}}$ & $1$ & $0.075$ & $0.7$ & $2$ & $2$ & $200$ \\
\end{tabular}
\label{tab: sim params 1}
\end{table}

\begin{figure}[ht]
\centering
\includegraphics[width=\columnwidth]{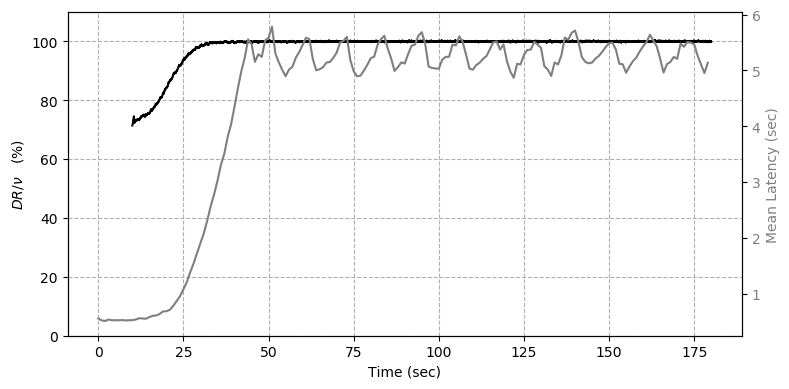}
\caption{Dissemination rate and mean latency for each node.}
\label{fig: dissemdelay 1}
\end{figure}

Figure \ref{fig: dissemdelay 1} shows the overall dissemination rate, $DR$, (see Definition \ref{def: dissemrate}) for this set of simulations alongside the mean latency (see Definition \ref{def: latency}) over all disseminated transactions. $DR$ is shown as a percentage of $\nu$ because $\nu$ is the maximum rate transactions can be disseminated. We observe that the dissemination rate converges to a value close to 100\% and that the mean latency converges to a steady state, in this case around 5 seconds. The mean latency depends heavily on the network diameter and delays associated with each hop, and the dissemination rate and latency convergence values depend on the rate setting parameters as  demonstrated below. \newline

\begin{figure}[ht]
\centering
\includegraphics[width=\columnwidth]{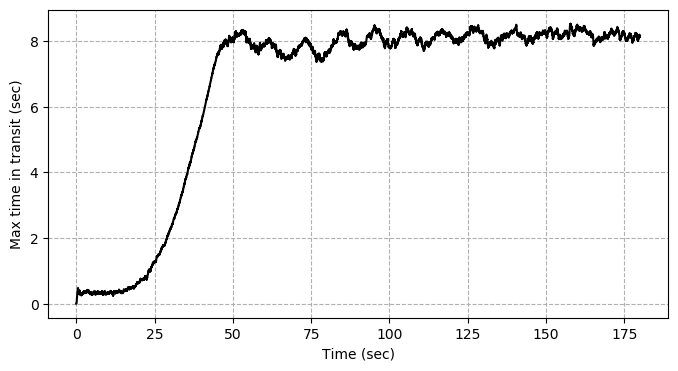}
\caption{Maximum time in transit, measured as time since issue for all undisseminated transactions, demonstrating that consistency is achieved.}
\label{fig: consistency 1}
\end{figure}

The consistency requirement is demonstrated by Figure \ref{fig: consistency 1} which shows the maximum time in transit. The time in transit can be measured for a transaction that is not yet disseminated, and is the time spent in the system, i.e., the time from when it was issued to the present. This value converges to a finite value in Figure \ref{fig: consistency 1},  demonstrating that consistency is achieved in the honest environment. Also note that no transactions were dropped by the buffer manager in these simulations, in line with research in \cite{suter1998design}. \newline

\begin{figure}[ht]
	\centering
	\includegraphics[width=\columnwidth]{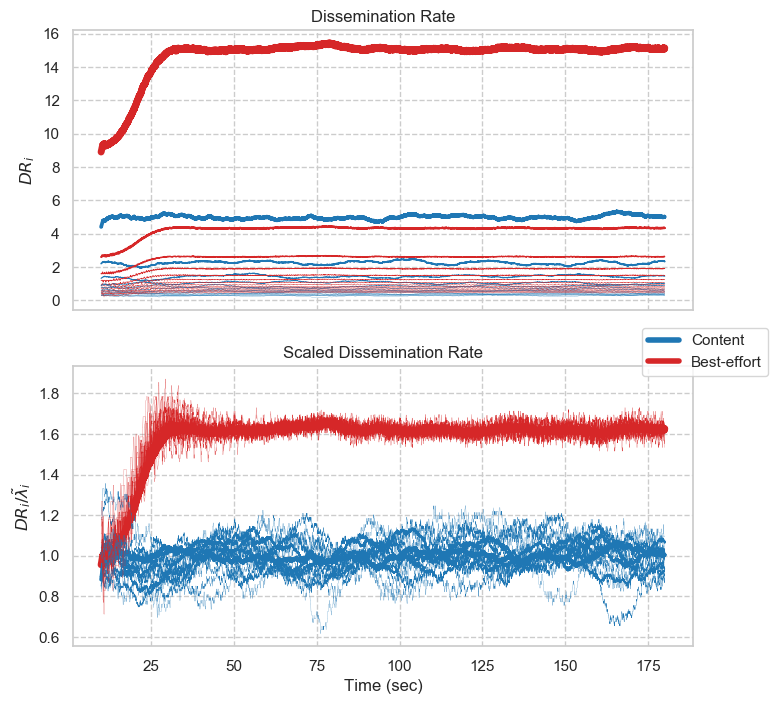}
	\caption{Dissemination rate and scaled dissemination rate of each node. The bottom plot of scaled dissemination rate demonstrates that fairness in dissemination rate is achieved.}
	\label{fig: ratefairness 1}
\end{figure}

Fairness in dissemination rate is demonstrated by Figures \ref{fig: ratefairness 1}. The upper subplots show the dissemination rate of each node, and the lower plot shows this dissemination rate scaled by each node's reputation. Best-effort nodes are plotted in red, and content nodes in blue, with the thickness of each trace proportional to the reputation of the relevant node, i.e., higher reputation nodes' rates are plotted with thicker lines. The bottom plot shows a particularly crucial result, namely it demonstrates that each node gets fair access to write to the ledger according to its reputation. In other words it implements Sybil protection, and we will show below that malicious actors can not tamper with this by deviating from the protocol.\newline

\begin{figure}[ht]
	\centering
	\includegraphics[width=\columnwidth]{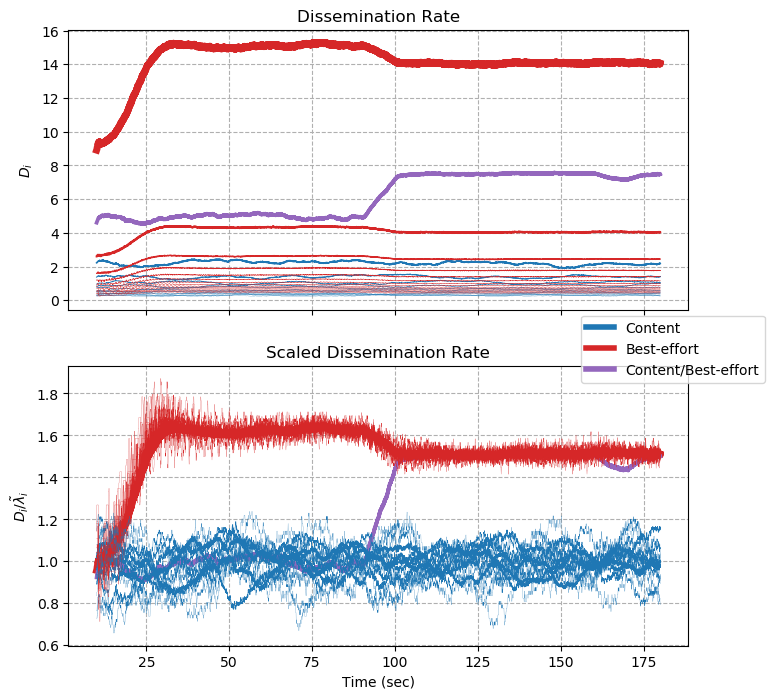}
	\caption{Dissemination rate and scaled dissemination rate of each node. The highest reputation content node (purple) switches to best-effort after 90 seconds and other best-effort nodes must adapt their rates.}
	\label{fig: ratefairness 1_1}
\end{figure}

{To further demonstrate fairness in dissemination rate, Figure \ref{fig: ratefairness 1_1} shows results from simulations in which the highest reputation content node switches to best-effort mode after 90 seconds. This node's dissemination rate is shown in purple and we can clearly see that all best-effort nodes adapt their issue rates to maintain fairness under the new traffic conditions.} \newline

\begin{figure}[ht]
\centering
\includegraphics[width=\columnwidth]{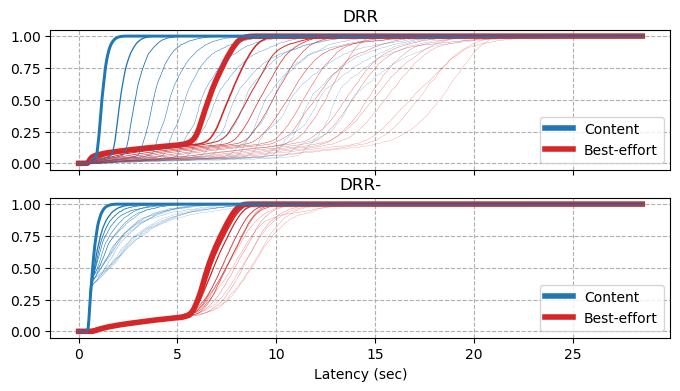}
\caption{Cumulative distribution of latency for each node for DRR scheduler and DRR$-$ scheduler. It is shown that only approximate fairness in latency is achieved, but DRR$-$ performs far better than standard DRR in this respect.}
\label{fig: latencyfairness 1}
\end{figure}

Fairness in latency is demonstrated by Figure \ref{fig: latencyfairness 1}, which shows the cumulative density function of latency across transactions issued by each node. Figure \ref{fig: latencyfairness 1} includes a comparison between the DRR$-$ scheduler and a standard DRR scheduler to demonstrate the improvements made by our design. The same convention for colour and line thickness as Figure \ref{fig: ratefairness 1} is used here. Clearly, fairness in latency is only approximately achieved, with lower reputation nodes experiencing higher latency than higher reputation nodes. However, our DRR$-$ scheduler a significant improvement over the standard DRR scheduler. Low reputation nodes receive slightly unfair treatment from the scheduler because their transactions are emptied more often  from inboxes since they have a lower issue rate and hence less backlog. In standard DRR, deficit can not be gained when an inbox is empty, so lower reputation nodes gain less deficit and inboxes become disproportionately backlogged with their transactions, resulting in higher delays. This problem is ameliorated in our scheduler (DRR$-$) by sometimes allowing deficit to be gained with an empty inbox.

\subsection{Malicious Environment}
\begin{figure}[ht]
\centering
\includegraphics[width=\columnwidth]{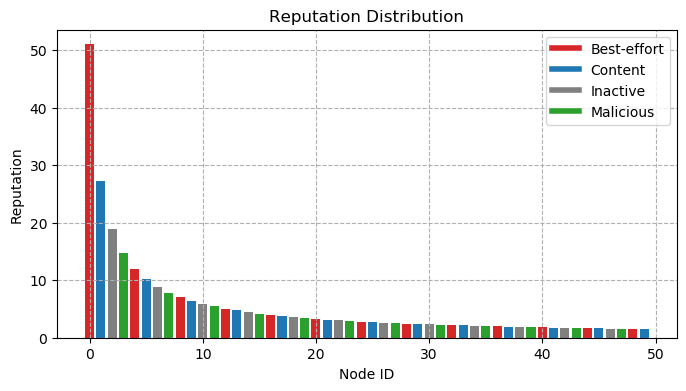}
\caption{Reputation distribution following a Zipf distribution with exponent 0.9. Nodes are Content, Best-effort, Inactive or Malicious as indicated by the colour of each bar.}
\label{fig: repdist 2}
\end{figure}

In order to test the security requirement, malicious nodes must be introduced to the simulation while consistency, fairness in dissemination rate and fairness in latency are not compromised for honest nodes. We focus our attention on the malicious behaviour defined in Definition \ref{def: malicious} as we can not anticipate all potential attack strategies. {This attack serves to show that a node can not simply inflate their dissemination rate beyond what its reputation allows.} Retaining the above network topology and reputation distribution, we introduce malicious nodes as illustrated in Figure \ref{fig: repdist 2}.  \newline

\begin{figure}[ht]
\centering
\includegraphics[width=\columnwidth]{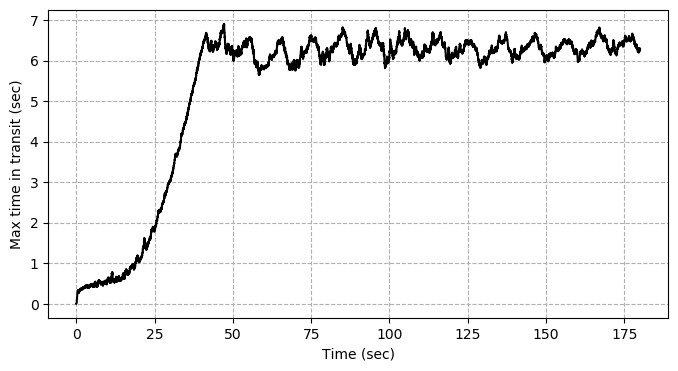}
\caption{Maximum time in transit for transactions issued by honest nodes, measured as time since issue for all undisseminated transactions, demonstrating that consistency is achieved.}
\label{fig: consistency 2}
\end{figure}

Figure \ref{fig: consistency 2} shows the maximum time spent in the system for transactions issued by honest nodes. This appears to converge  as before, indicating that consistency is still achieved. \newline

\begin{figure}[ht]
\centering
\includegraphics[width=\columnwidth]{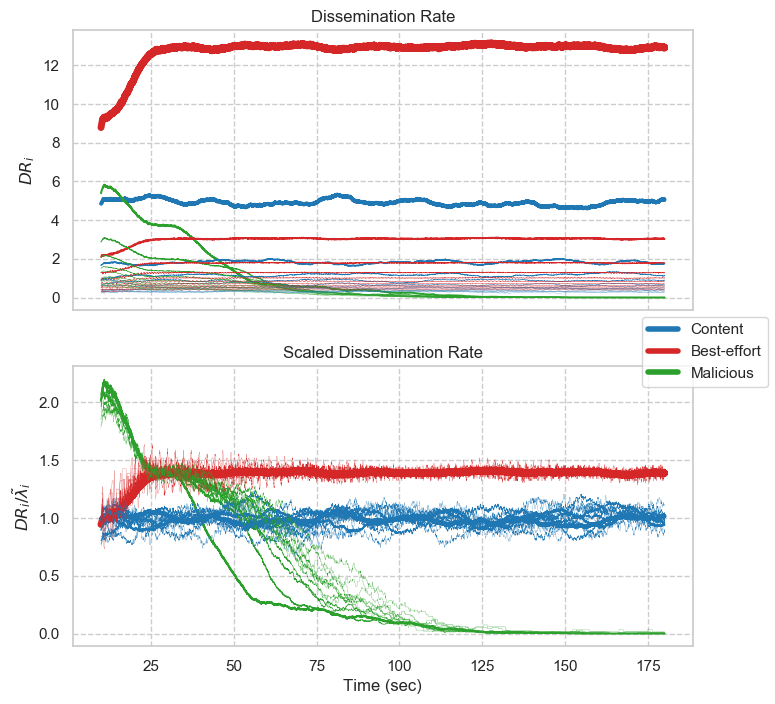}
\caption{Dissemination rate and scaled dissemination rate for each node. The bottom plot of scaled dissemination rate demonstrates that fairness in dissemination rate is achieved for honest nodes, while malicious nodes are penalised by the buffer management and experience lower dissemination rates.}
\label{fig: ratefairness 2}
\end{figure}

Figure \ref{fig: ratefairness 2} shows the dissemination rates and scaled dissemination rates for this set of simulations. Clearly fairness is still achieved for honest nodes. The dissemination rate of the malicious nodes initially begins to converge to the max-min fair value also.  But when they begin to cause excessive congestion, the buffer management component begins dropping malicious transactions,  and the malicious nodes' dissemination rate falls to zero. \newline

\begin{figure}[ht]
\centering
\includegraphics[width=\columnwidth]{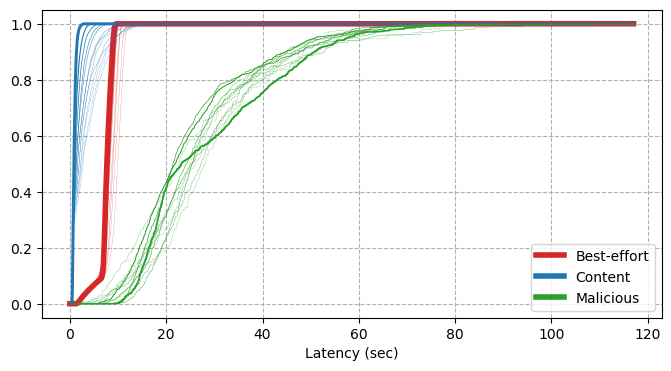}
\caption{Cumulative distribution of latency for each node. Malicious nodes are shown to experience higher latency, while approximate fairness in latency is retained for honest nodes.}
\label{fig: latencyfairness 2}
\end{figure}

Figure \ref{fig: latencyfairness 2} demonstrates that approximate latency fairness is still achieved for the honest nodes but that malicious nodes experience far higher latency. Only the DRR$-$ scheduler is displayed in this case as there is no further need to compare with standard DRR as in Figure \ref{fig: latencyfairness 1}.

\subsection{Sensitivity Analysis}
{We now demonstrate how the network responds to tuning of the rate setting parameters. We focus particularly on the increase parameter $A$, the decrease parameter $\beta$, the work threshold $W$, and the total number of nodes $|/mathcal{M}|$.} The wait time $\tau$ should simply be chosen long enough so that successive decreases do not falsely occur after a congestion event. \newline

\begin{figure}[ht]
\centering
\includegraphics[width=\columnwidth]{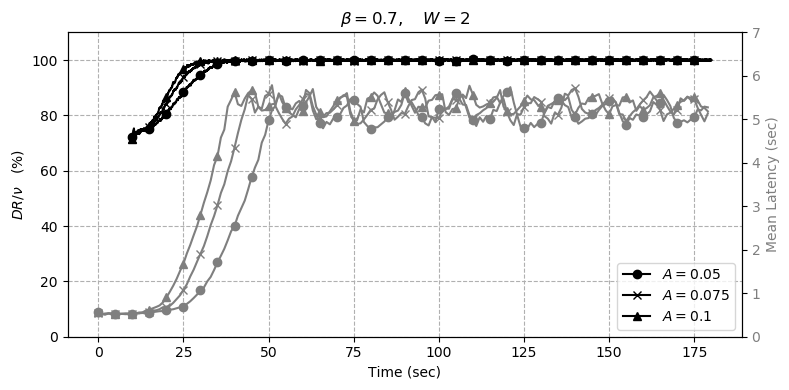}
\caption{Combined dissemination rate as a percentage of $\nu$, and mean latency, changing the additive increase parameter $A$.}
\label{fig: ratesettercomp 1}
\end{figure}

First, consider the increase parameter $A$. Beginning with the simulation parameters given in Table \ref{tab: sim params 1}, we demonstrate the effect of increasing and decreasing $A$ in Figure \ref{fig: ratesettercomp 1}. It is shown that increasing $A$ results in faster convergence to the equilibrium dissemination rate and mean latency, but that it has little impact on the equilibrium expected values. \newline

\begin{figure}[ht]
\centering
\includegraphics[width=\columnwidth]{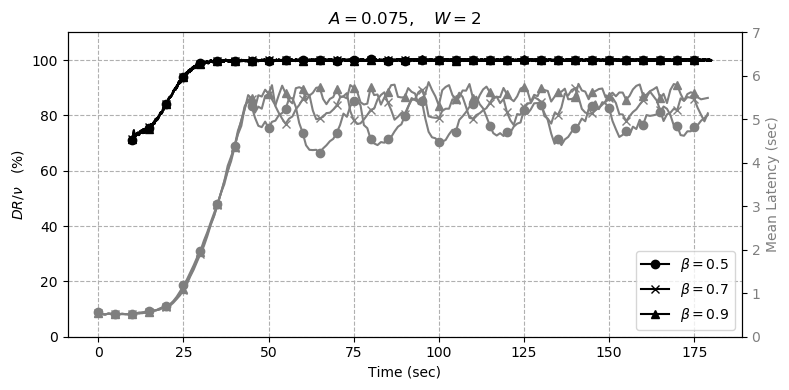}
\caption{Combined dissemination rate as a percentage of $\nu$, and mean latency, changing the multiplicative decrease parameter $\beta$.}
\label{fig: ratesettercomp 2}
\end{figure}

Next, we analyse the impact of the decrease parameter $\beta$. Similar to the last experiment, we adjust $\beta$ while fixing the other simulation parameters listed in Table \ref{tab: sim params 1}. Figure \ref{fig: ratesettercomp 2} shows the corresponding results. A lower decrease parameter causes the issue rate to more drastically decrease at congestion events. Thus we observe more oscillation in the mean latency, indicating that queue lengths oscillate more. Also we observe a very slight decrease in dissemination rate as $\beta$ is decreased. \newline

\begin{figure}[ht]
\centering
\includegraphics[width=\columnwidth]{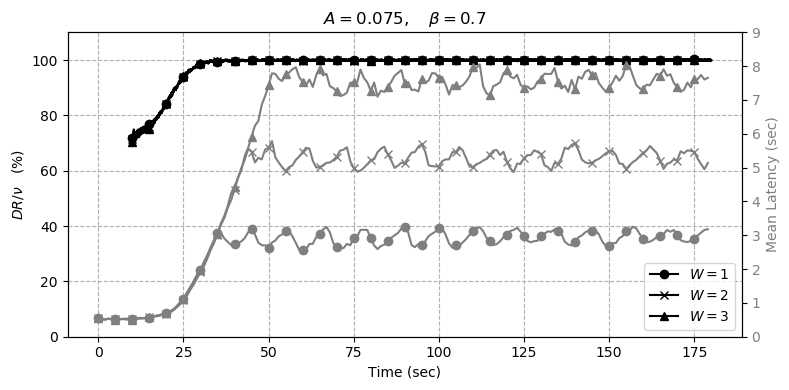}
\caption{Combined dissemination rate as a percentage of $\nu$, and mean latency, changing the work threshold parameter $W$.}
\label{fig: ratesettercomp 3}
\end{figure}

Finally, we demonstrate the impact of the work threshold $W$. This threshold corresponds to the level of backlog which is considered congestion, so this parameter has a significant impact on the queue lengths and hence the latency. This is illustrated in Figure \ref{fig: ratesettercomp 3}, in which $W$ is changed while keeping all other simulation parameters from Table \ref{tab: sim params 1} fixed. We observe that there is a very clear impact on the mean latency resulting from this choice, and less so from different choices of $A$ and $\beta$. The significance of $W$ is directly linked to the equilibrium level of congestion at which the rate setting responds, while $A$ and $\beta$ primarily determine how quickly the system reaches that equilibrium, and how aggressively the rate setting reacts to each congestion event, respectively. Note that if we decrease the threshold $W$ too much, noise in the measurements of the inbox can become significant and fairness of the rate setting can be compromised. The lower limit of $W$ to minimise latency while preserving fairness should be determined experimentally for deployment in a real network. \newline

\begin{figure}[ht]
\centering
\includegraphics[width=\columnwidth]{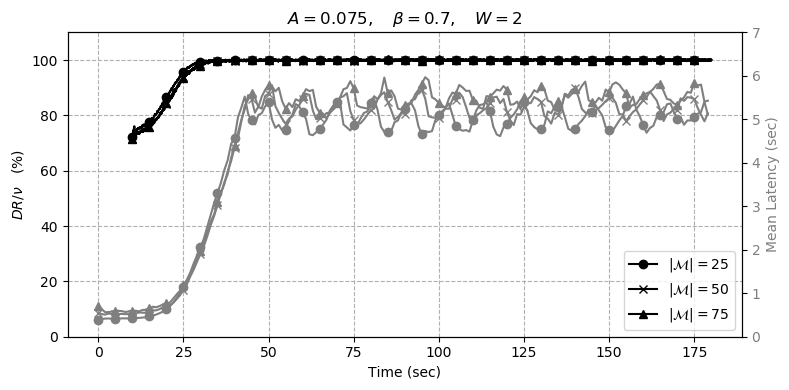}
\caption{Combined dissemination rate as a percentage of $\nu$, and mean latency with varying number of nodes $|\mathcal{M}|$ in the network.}
\label{fig: ratesettercomp m}
\end{figure}

{The rate setting parameters here are designed to scale with the network, so new nodes joining  should not require re-tuning of parameters. Figure \ref{fig: ratesettercomp m} presents dissemination rate and delay as the number of nodes is changed, with all other access control algorithm parameters remaining unchanged.} The total reputation as given in Figure \ref{fig: repdist 1} is conserved, but is redistributed among nodes to retain the Zipf distribution as nodes join and leave the network. It is clear that the algorithm performs well without requiring retuning of parameters as the network scales, although delay is increased slightly as the diameter of the network increases. \newline

Future work will see our algorithm deployed in a real DLT network (IOTA's GoShimmer network) with varying number of nodes, real transaction traffic and a range of network topologies which will allow us to further verify the robustness of our parameter choices. Parameters relating to inbox lengths and scheduler parameters will require consideration of the buffer capacity available to nodes and the number of nodes in the network, which will also be examined when the algorithm is deployed in our test network.

\subsection{IoT Devices and Variable Transaction Work}
For simplicity of presentation, all transactions in the simulations presented so far require equal work to write and are therefore treated equally by the access control. In the IoT setting, transactions may contain sensor readings, machine-to-machine micro-payments, or a range of application-specific data, all of which  requiring different levels of work to write. As opposed to  IoT data transactions, we expect that transactions which transfer currency will typically require more work due to processes such as validity checks and updating account balances. In the following set of simulations, we demonstrate that our algorithm can handle variable transaction work requirements, and  that lower work transactions actually experience lower latency, giving an advantage to IoT devices issuing such transactions. \newline

In order to illustrate the impact of varying transaction work, consider two representative node types:
\begin{itemize}
\item \emph{Value node}: all transactions are value transfers and require one unit of work.
\item \emph{IoT node}: transactions contain an array of data types and require random work uniformly distributed on the interval $[0.25, 0.75]$.
\end{itemize}
Consider the same reputation distribution and operation modes as illustrated in Figure \ref{fig: repdist 1} with all even numbered nodes as value nodes and all odd number nodes as IoT nodes (note that nodes are numbered from 0 to 49). The access control parameters are as given in Table \ref{tab: sim params 1}. \newline

\begin{figure}[ht]
\centering
\includegraphics[width=\columnwidth]{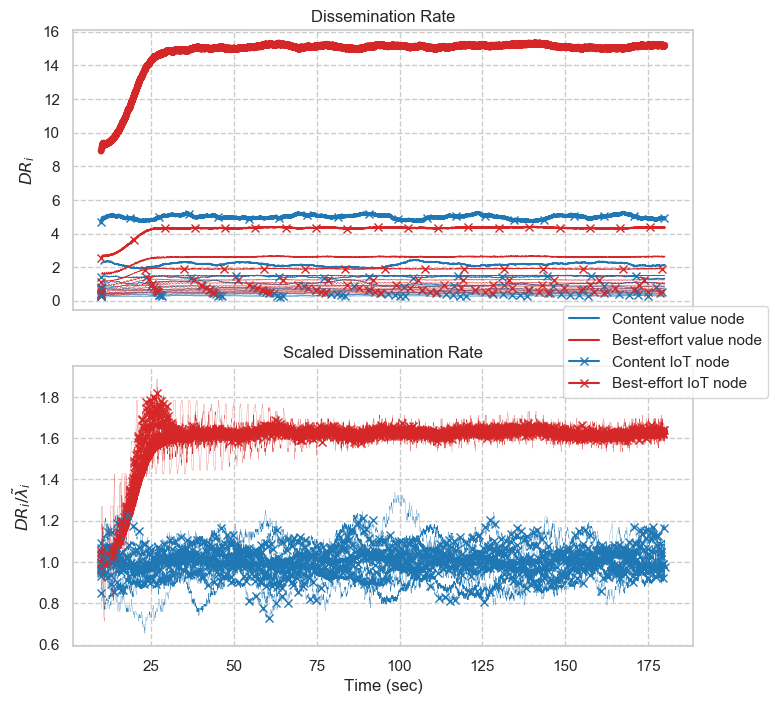}
\caption{Dissemination rates of each node with a mixture IoT nodes and value nodes. The bottom plot demonstrates that fairness in dissemination rate is achieved in the presence of variable transaction work requirements.}
\label{fig: ratefairness iot}
\end{figure}

\begin{figure}[ht]
\centering
\includegraphics[width=\columnwidth]{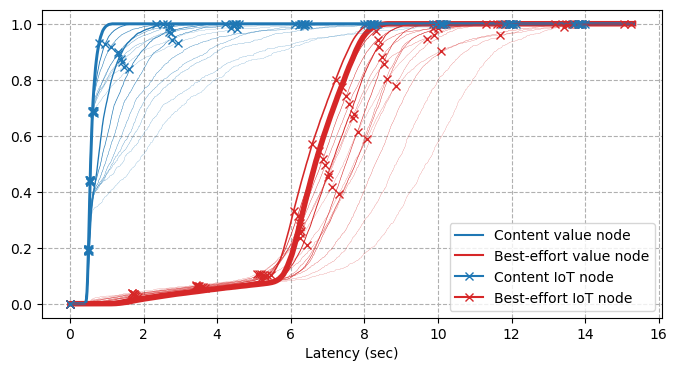}
\caption{Cumulative distribution of latency for each node with a combination of IoT nodes and value nodes. It is clear that approximate fairness in latency is still achieved in the presence of variable transaction work requirements.}
\label{fig: latencyfairness iot}
\end{figure}

Figure \ref{fig: ratefairness iot} shows that fairness is still achieved with this combination of IoT nodes and value nodes. Figure \ref{fig: latencyfairness iot} shows the latency for each node. Approximate fairness in latency is still achieved, and more notably, IoT nodes have lower latency compared with the equivalent nodes in earlier simulations.

\subsection{Comparison to Proof of Work Access Control}
{At the time of writing, PoW access control is the state-of-the-art for DAG-based DLTs.} PoW access control for DAGs is very straightforward: the difficulty of the PoW is set for the protocol and this determines how rapidly a node can create transactions with a valid PoW puzzle solved, and transactions are simply scheduled in FIFO order. The PoW difficulty should be set such that writing resources are well utilised, but nodes with scarce writing resources can write all incoming transactions to their ledger without becoming congested. However, if the PoW difficulty is set too high, writing resources will be underutilised and dissemination rate will not be maximised. On the other hand, if it set too low, transactions could be issued too rapidly for nodes with low writing power, and they will become overwhelmed. \newline

For the purpose of illustration, suppose we estimate the combined computing power of all nodes in the network, and we set the PoW difficulty such that if all nodes were active at the same time, the nodes with the lowest writing power would just be able to keep up. We then have three cases which can arise:
\begin{enumerate}
\item Some nodes are inactive, or the active computing power is lower than estimated.
\item The estimate of computing power matches the active computing power in the network.
\item The active computing power in the network is higher than estimated, or new nodes have recently joined with additional computing power above the estimated level.
\end{enumerate} 

\begin{figure}[ht]
\centering
\includegraphics[width=\columnwidth]{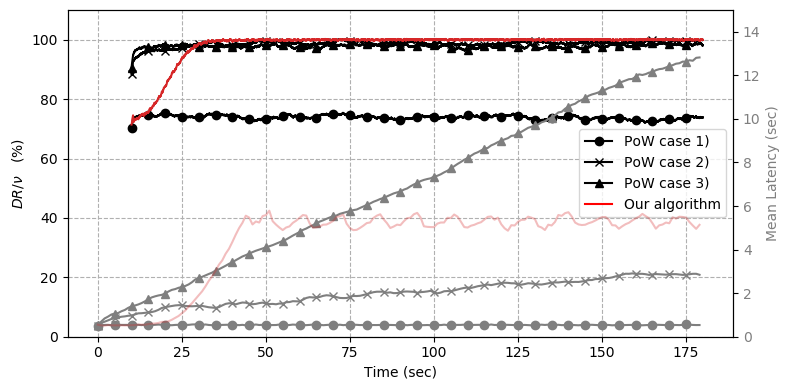}
\caption{{Dissemination rate as a percentage of maximum scheduling rate, $\nu$, and mean latency for cases 1)--3) of PoW access control, shown alongside our algorithm with parameters given in Table \ref{tab: sim params 1}.}}
\label{fig: dissemdelay powcomp}
\end{figure}

{Cases 1)--3) are compared to our algorithm in Figure \ref{fig: dissemdelay powcomp}}. We simulate these cases using the network from Section \ref{subsec: honest}, with computing power distribution as per the Zipf distribution of reputation given in this section. We set the PoW difficulty such that all nodes using their full computing power could issue transactions at a rate $\nu$. For case 1), the inactive nodes as shown in Figure \ref{fig: repdist 1} are inactive, and all others use their full computing power, resulting in lower active computing power than estimated. For case 2), we take all nodes as active and utilising their full computing power, which results in the active computing power precisely matching the estimate. For case 3), we increase each nodes computing power by 5\%, resulting in higher active computing power than estimated.

{From Figure \ref{fig: dissemdelay powcomp} we see that: in case 1) latency is low but resources are underutilised as is evident from the dissemination rate below 80\%; in case 2) the utilisation is good, but the latency gradually increases as some queues build up at the slowest nodes; in case 3) the utilisation is high but latency blows up due to nodes with low resources being unable to keep up with the rate of transactions being issued; with our algorithm we acheive high utilisation and maintain a stable latency as nodes can all manage their queues.} \newline

It is clear that PoW is very sensitive to the difficulty setting, performing poorly or failing completely if this is estimated incorrectly. This is a major issue because techniques for adapting the difficulty of PoW based on the estimated active computing power, such as that used in the Bitcoin network \cite{nakamoto2008bitcoin}, operate over excessively long time scales of many hours or even days. In our access control algorithm, on the other hand, nodes do not need to waste energy on solving PoW puzzles and can adapt their issue rate immediately in response to traffic observed in their inbox. \newline

\textbf{Remark:} the results for PoW access control presented above demonstrate just some of the weaknesses of existing access control for DAG-based DLTs. Case 2) may seem almost acceptable, but note that this is not even practically achievable and that all three PoW cases correspond to hugely wasteful energy consumption and performance limitation, and the reality is that PoW is a legacy access control mechanism which prevents mainstream adoption of DLT in the IoT setting. This work offers the first IoT-friendly alternative.

\section{Conclusions}
\label{sec: conclusions}
We have presented an access control algorithm for DAG-based distributed ledgers which enforces a resource allocation to nodes based on their reputation. Our solution is especially suitable for the IoT setting because nodes are not required to wastefully commit computing resources in order to contribute to the ledger, as is the case in traditional DLTs. Additionally, the DAG-based ledger structure permits high transaction dissemination rate and low latency because transactions are not limited to being added in blocks. \newline

{
The main features of our access control solution are an efficient fair scheduler, a TCP-inspired rate setting algorithm, and a buffer management scheme. We have shown, via network simulations, that our algorithm:
\begin{itemize}
	\item ensures resources are allocated fairly and securely. The algorithm permits high utilisation of resources while preventing excessive congestion which could otherwise cause large delays and buffer overflows;
	\item is resilient against malicious agents wishing to claim more than their fair share of network resources;
	\item is robust to changes in parameters;
	\item permits a range of transaction types and sizes which may require differentiated treatment, making our algorithm applicable to networks with mixed node types, such as IoT nodes storing sensor data on the ledger, and larger devices making financial transactions;
	\item improves significantly on the state of the art in access control, namely PoW. Our comparison demonstrate PoW's shortcomings and further motivates our new approach.
\end{itemize}
}

At the time of writing, our algorithm is being tested on IOTA Foundation's GoShimmer test network after which it will be deployed in IOTA's main network, forming the security backbone of the IOTA ledger. We also expect that this work will generate great interest from the DLT and networking communities alike, and that the algorithm presented here will inspire many others like it, enabling a new generation of IoT-friendly distributed ledgers. {We expect that particular research efforts will focus on mathematical analysis of algorithms like ours and on further improving scalability to accommodate the growing number of IoT devices using DLTs.}

\bibliographystyle{IEEEtran}
\bibliography{refs}

\begin{IEEEbiography}[{\includegraphics[width=1in,height=1.25in,clip,keepaspectratio]{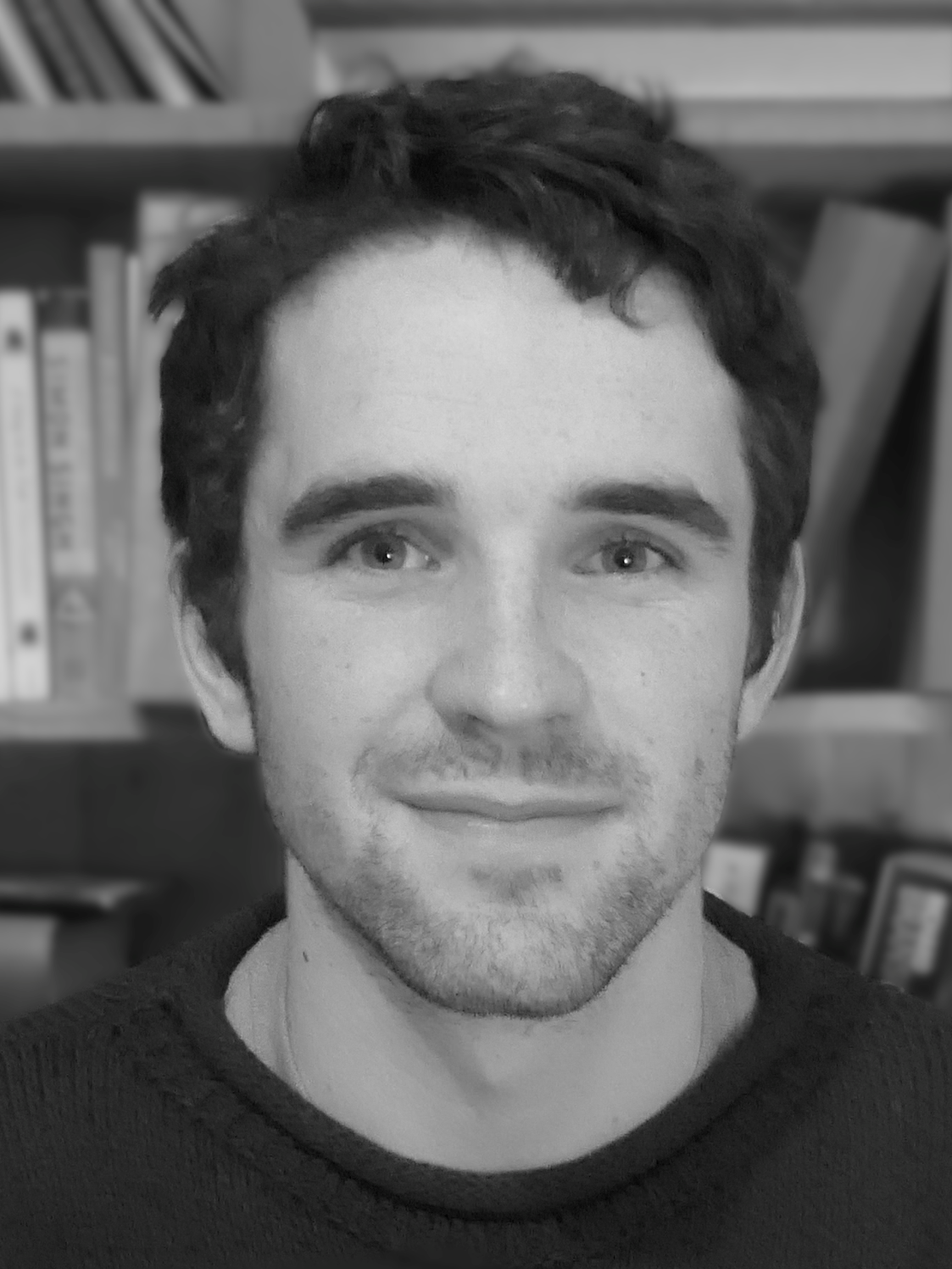}}]{Andrew Cullen}
	is an IBM PhD Fellow at the Dyson School of Design Engineering at Imperial College London, United Kingdom. He received the B.Sc. and M.E. from University College Dublin in 2016 and 2018 respectively. His research is focused on trust in distributed systems, with a focus on IoT, cyber-physical systems and human-machine interactions. He has collaborated with IOTA Foundation's networking research team to develop the next generation of their public, IoT-focused distributed ledger.
\end{IEEEbiography}

\begin{IEEEbiography}[{\includegraphics[width=1in,height=1.25in,clip,keepaspectratio]{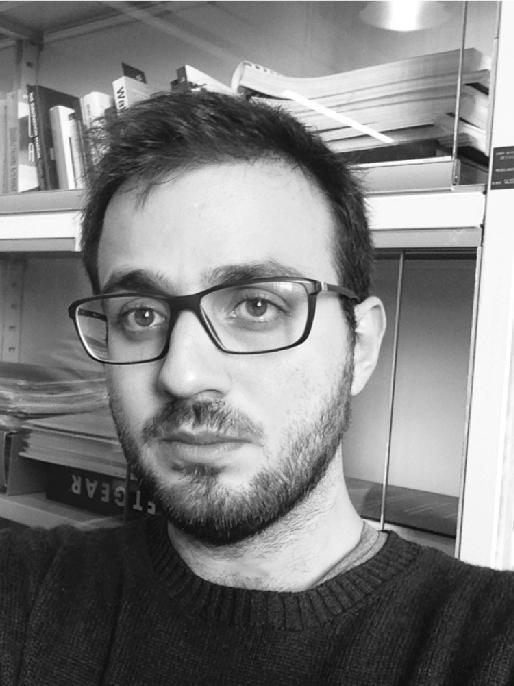}}]{Pietro Ferraro}
	received the Ph.D. in control and electrical engineering from the University of Pisa, Italy, in 2018. He is currently a Research Associate at the Dyson School of Design Engineering, Imperial College London and a Research Scientist at the IOTA Foundation. His research interests include control theory, distributed ledger technologies and the sharing economy. He works closely with a number of research teams within IOTA, primarily on topics related to networking and compliance. 
\end{IEEEbiography}

\begin{IEEEbiography}[{\includegraphics[width=1in,height=1.25in,clip,keepaspectratio]{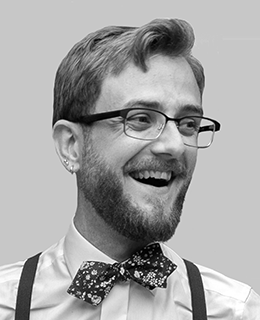}}]{William Sanders}
	is Director of Research at IOTA Foundation. He received the Ph.D. in Mathematics from the University of Kansas in 2015 and conducted postdoctoral research at Norwegian Institute of Science and Technology from 2015 to 2018. He has expertise in commutative algebra, category theory and combinatorics, and much of his research within IOTA has focused on developing a unified theory of distributed ledgers. He has contributed to a wide range of projects within IOTA including networking problems, protocol improvements, consensus questions, reputation systems and autopeering.
\end{IEEEbiography}

\begin{IEEEbiography}[{\includegraphics[width=1in,height=1.25in,clip,keepaspectratio]{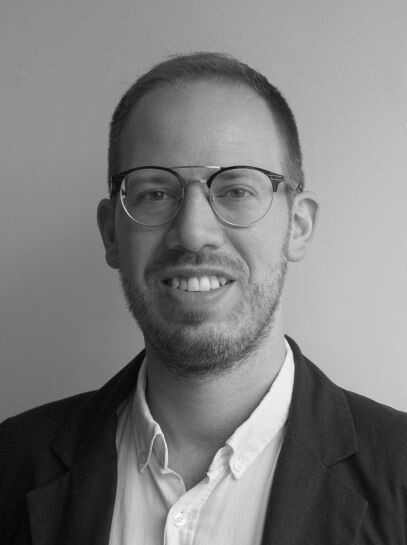}}]{Luigi Vigneri}
	is a senior research scientist at IOTA Foundation since 2018, leading the networking team. His current research interests concern network optimisation and scalability in the context of distributed ledger technologies. Prior to IOTA, Luigi joined Huawei Research Lab in Paris as a postdoctoral researcher working on 5G networks, online machine learning and artificial intelligence. In 2017, he obtained his Ph.D. in Mobile Communications from EURECOM, France, with a thesis on caching popular content in vehicular networks.
\end{IEEEbiography}

\begin{IEEEbiography}[{\includegraphics[width=1in,height=1.25in,clip,keepaspectratio]{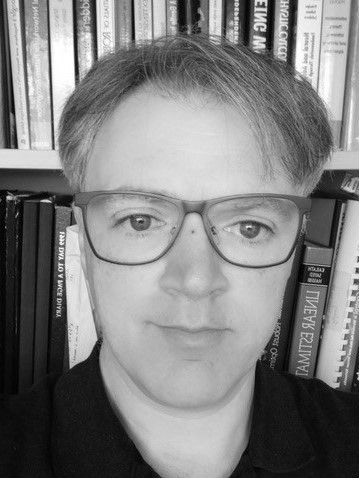}}]{Robert Shorten}
	 holds the Chair of Cyber-physical Systems Design in the Dyson School of Design Engineering Imperial College London. He has been employed by IBM Research (lead- ing the Control and Optimisation activities in the Smart Cities Research Centre), by Daimler-Benz Research, and by AEG-Westinghouse, as well as holding appointments at TU Berlin, University College Dublin, and Maynooth University. His work has been applied in multiple domains (net- working, automotive, smart cities), and has been implemented in production systems (automotive, Linux and FreeBSD, distributed ledgers).  He is a co-author of AIMD Dynamics and Distributed Resource Allocation (SIAM, 2016), Electric and Plug-in Vehicle Networks: Optimisation and Control (CRC Press 2017) and an editor of Analytics for the Sharing Economy: Mathematics, Engineering and Business Perspectives (Springer 2020). Professor Shorten is Chair of the Board of Directors of Future Mobility Campus Ireland, Global Technical Program Co-Chair of ICCVE 2022, and Industry Chair of ECC, 2022.
\end{IEEEbiography}

\end{document}